# Spin-orbit coupling in the hydrogen atom, the Thomas precession, and the exact solution of Dirac's equation

Masud Mansuripur
College of Optical Sciences, The University of Arizona, Tucson, Arizona 85721

**Abstract**. Bohr's model of the hydrogen atom can be extended to account for the observed spin-orbit interaction, either with the introduction of the Thomas precession,[1] or with the stipulation that, during a spin-flip transition, the orbital radius remains intact.[2] In other words, if there is a desire to extend Bohr's model to accommodate the spin of the electron, then experimental observations mandate the existence of the Thomas precession, which is a questionable hypothesis,[2] or the existence of artificially robust orbits during spin-flip transitions. This is tantamount to admitting that Bohr's model, which is a poor man's way of understanding the hydrogen atom, is of limited value, and that one should really rely on Dirac's equation for the physical meaning of spin, for the mechanism that gives rise to the gyromagnetic coefficient $g = 2$, for Zeeman splitting, for relativistic corrections to Schrödinger's equation, for Darwin's term, and for the correct ½ factor in the spin-orbit coupling energy.

**1. Introduction**. This paper is a tutorial presentation of the standard Dirac equation of quantum mechanics,[3] followed by a rigorous solution of the equation for the hydrogen atom.[3-5] In Sec.2 we introduce the Dirac equation and show its connections to the Schrödinger equation and to Pauli's spin matrices. Here, we also encounter the Darwin Hamiltonian, the gyromagnetic coefficient (the so-called *g*-factor) associated with the spin magnetic moment of the electron in the context of the Zeeman effect, and a term responsible for the electron's spin-orbit coupling in the presence of the electric charge of the nucleon(s). Section 3 is devoted to solving Dirac's equation for hydrogen-like atoms. Considering the central role that angular momentum operators play in the solution of Dirac's equation, the general concepts and operations pertaining to a quantum-mechanical treatment of spin and orbital angular momenta are discussed in an Appendix at the end of the paper, where, all the formulas needed for a thorough understanding of the solution of Dirac's equation are derived. Section 4 examines the radial dependence of the wave-function of the electron in a hydrogen-like atom, and arrives at closed form expressions for the relevant functions. After listing some useful identities in Sec.5, we derive an explicit formula in Sec.6 for the coefficients that appear in the radial component of the electron's wave-function. A procedure for normalizing the wave-function is outlined in Sec.7, and a few examples of calculations of eigen-energies and eigen-functions are provided in Sec.8. A brief summary and concluding remarks form the subject of the final section.

**2. The Dirac equation**. Dirac's Hamiltonian operator is $\widehat{H} = -\mathrm{i}\hbar c(\boldsymbol{\alpha} \cdot \boldsymbol{\nabla}) + mc^2 \beta$, where $\boldsymbol{\alpha} = \alpha_x \widehat{\boldsymbol{x}} + \alpha_y \widehat{\boldsymbol{y}} + \alpha_z \widehat{\boldsymbol{z}}$ is a $4 \times 4$ vector matrix, $\beta$ is a $4 \times 4$ scalar matrix, $-\mathrm{i}\hbar \boldsymbol{\nabla} = -\mathrm{i}\hbar(\partial_x \widehat{\boldsymbol{x}} + \partial_y \widehat{\boldsymbol{y}} + \partial_z \widehat{\boldsymbol{z}})$ is the (canonical) momentum operator, and $mc^2$ is the rest-energy of the electron. Note that $\widehat{\boldsymbol{x}}$, $\widehat{\boldsymbol{y}}$, and $\widehat{\boldsymbol{z}}$ are units-vectors along the Cartesian coordinate axes, not operators. In Dirac's own notation, the $4 \times 4$ matrices $\alpha_x, \alpha_y, \alpha_z, \beta$ are given in terms of the Pauli matrices $\sigma_x, \sigma_y, \sigma_z$ and the $2 \times 2$ identity matrix $\mathbb{I}$, as follows:

$$\alpha_x = \begin{pmatrix} 0 & \sigma_x \\ \sigma_x & 0 \end{pmatrix}, \quad \text{where } \sigma_x = \begin{pmatrix} 0 & 1 \\ 1 & 0 \end{pmatrix}; \tag{1a}$$

$$\alpha_y = \begin{pmatrix} 0 & \sigma_y \\ \sigma_y & 0 \end{pmatrix}, \quad \text{where } \sigma_y = \begin{pmatrix} 0 & -\mathrm{i} \\ \mathrm{i} & 0 \end{pmatrix}; \tag{1b}$$

$$\alpha_z = \begin{pmatrix} 0 & \sigma_z \\ \sigma_z & 0 \end{pmatrix}, \quad \text{where } \sigma_z = \begin{pmatrix} 1 & 0 \\ 0 & -1 \end{pmatrix}; \tag{1c}$$

$$\beta = \begin{pmatrix} \mathbb{I} & 0 \\ 0 & -\mathbb{I} \end{pmatrix}, \quad \text{where } \mathbb{I} = \begin{pmatrix} 1 & 0 \\ 0 & 1 \end{pmatrix}. \tag{1d}$$

The Dirac equation may thus be written

$$i\hbar \partial_t \psi(\boldsymbol{r},t) = -i\hbar c(\alpha_x \partial_x + \alpha_y \partial_y + \alpha_z \partial_z)\psi(\boldsymbol{r},t) + mc^2 \beta \psi(\boldsymbol{r},t). \tag{2}$$

Here, the wave-function $\psi(\boldsymbol{r},t)$, representing the state $|\psi\rangle$ of the electron, is Dirac's 4-component spinor, namely,

$$\psi(\boldsymbol{r},t) = |\psi\rangle = \begin{bmatrix} \psi_{+\uparrow}(\boldsymbol{r},t) \\ \psi_{+\downarrow}(\boldsymbol{r},t) \\ \psi_{-\uparrow}(\boldsymbol{r},t) \\ \psi_{-\downarrow}(\boldsymbol{r},t) \end{bmatrix}. \tag{3}$$

In the above notation, the $\pm\uparrow\downarrow$ subscripts distinguish the *dominant* components of the spinor in conjunction with positive-energy–spin-up, positive-energy–spin-down, negative-energy–spin-up, and negative-energy–spin-down states of the electron. In the presence of an external electromagnetic field, Dirac's equation incorporates the field's scalar and vector potentials, $\Phi(\boldsymbol{r},t)$ and $\boldsymbol{A}(\boldsymbol{r},t)$, in the so-called minimal coupling scheme, as follows:

$$(i\hbar \partial_t - e\Phi)\psi = c\boldsymbol{\alpha}\cdot(-i\hbar\boldsymbol{\nabla} - e\boldsymbol{A})\psi + mc^2\beta\psi. \tag{4}$$

Here, $e$ (a negative entity) is the charge of an electron. The above equation may be broken down into two coupled differential equations for two-component spinors $\psi_+(\boldsymbol{r},t)$ and $\psi_-(\boldsymbol{r},t)$. The resulting equations are

$$(i\hbar \partial_t - e\Phi - mc^2)\psi_+ = c\boldsymbol{\sigma}\cdot(-i\hbar\boldsymbol{\nabla} - e\boldsymbol{A})\psi_-, \tag{5a}$$

$$(i\hbar \partial_t - e\Phi + mc^2)\psi_- = c\boldsymbol{\sigma}\cdot(-i\hbar\boldsymbol{\nabla} - e\boldsymbol{A})\psi_+. \tag{5b}$$

For static potentials $\boldsymbol{A}(\boldsymbol{r})$ and $\Phi(\boldsymbol{r})$, the eigenstate $\psi(\boldsymbol{r},t) = \exp(-i\mathcal{E}t/\hbar)\psi(\boldsymbol{r})$ of the Dirac Hamiltonian with a positive eigen-energy $\mathcal{E} = mc^2 + \tilde{\mathcal{E}}$, when placed in Eq.(5b), yields $\psi_-(\boldsymbol{r})$ as follows:

$$\psi_- = c(2mc^2 + \tilde{\mathcal{E}} - e\Phi)^{-1}\boldsymbol{\sigma}\cdot(-i\hbar\boldsymbol{\nabla} - e\boldsymbol{A})\psi_+$$

$$\cong -\frac{i\hbar}{2mc}\left(1 - \frac{\tilde{\mathcal{E}} - e\Phi}{2mc^2}\right)\boldsymbol{\sigma}\cdot(\boldsymbol{\nabla} - ie\boldsymbol{A}/\hbar)\psi_+. \tag{6}$$

Substitution for $\psi_-$ from Eq.(6) into Eq.(5a) now leads to the following equation for $\psi_+$:

$$(\tilde{\mathcal{E}} - e\Phi)\psi_+ = -\frac{\hbar^2}{2m}\boldsymbol{\sigma}\cdot(\boldsymbol{\nabla} - ie\boldsymbol{A}/\hbar)\left(1 - \frac{\tilde{\mathcal{E}} - e\Phi}{2mc^2}\right)\boldsymbol{\sigma}\cdot(\boldsymbol{\nabla} - ie\boldsymbol{A}/\hbar)\psi_+. \tag{7}$$

Considering that Pauli's matrices $\sigma_x, \sigma_y, \sigma_z$ can readily slip through the operators, the following identities may be used to simplify the preceding equation:

$$\sigma_x \sigma_x = \sigma_y \sigma_y = \sigma_z \sigma_z = \mathbb{I}, \tag{8a}$$

$$\sigma_x \sigma_y = -\sigma_y \sigma_x = i\sigma_z, \tag{8b}$$

$$\sigma_y \sigma_z = -\sigma_z \sigma_y = i\sigma_x, \tag{8c}$$

$$\sigma_z \sigma_x = -\sigma_x \sigma_z = i\sigma_y. \tag{8d}$$

Ignoring for the moment the coefficient $ie/\hbar$ of $\boldsymbol{A}$, and also denoting $\left(1 - \frac{\tilde{\mathcal{E}} - e\Phi}{2mc^2}\right)$ by $\varphi$, we write the operator appearing on the right-hand-side of Eq.(7) as follows:

$$[\sigma_x(\partial_x - A_x) + \sigma_y(\partial_y - A_y) + \sigma_z(\partial_z - A_z)]\varphi[\sigma_x(\partial_x - A_x) + \sigma_y(\partial_y - A_y) + \sigma_z(\partial_z - A_z)]$$

$$= (\partial_x - A_x)\varphi(\partial_x - A_x) + (\partial_y - A_y)\varphi(\partial_y - A_y) + (\partial_z - A_z)\varphi(\partial_z - A_z)$$

$$+ i\sigma_x[(\partial_y - A_y)\varphi(\partial_z - A_z) - (\partial_z - A_z)\varphi(\partial_y - A_y)]$$

$$+ i\sigma_y[(\partial_z - A_z)\varphi(\partial_x - A_x) - (\partial_x - A_x)\varphi(\partial_z - A_z)]$$

$$+ i\sigma_z[(\partial_x - A_x)\varphi(\partial_y - A_y) - (\partial_y - A_y)\varphi(\partial_x - A_x)]$$

Use the following vector identities:
$\nabla \cdot (A\psi) = A \cdot \nabla\psi + (\nabla \cdot A)\psi$
$\nabla \times (A\psi) = (\nabla \times A)\psi - A \times \nabla\psi$

$$= \varphi\nabla^2 + \nabla\varphi \cdot (\nabla - A) - \varphi(\nabla \cdot A) - 2\varphi A \cdot \nabla + \varphi A \cdot A + i\sigma \cdot (\nabla\varphi \times \nabla + A \times \nabla\varphi - \varphi\nabla \times A). \tag{9}$$

Putting everything back into Eq.(7), we will have

$$\tilde{\mathcal{E}}\psi_+ = \left\{\left(1 - \frac{\tilde{\mathcal{E}} - e\Phi}{2mc^2}\right)\left[-\frac{\hbar^2}{2m}\nabla^2 + i\frac{e\hbar}{m}A \cdot \nabla + i\frac{e\hbar}{2m}\nabla \cdot A + \frac{e^2}{2m}A \cdot A - \frac{e\hbar}{2m}\sigma \cdot B\right] + e\Phi$$

$$- i\frac{e\hbar}{4m^2c^2}\nabla\Phi \cdot (-i\hbar\nabla - eA) + \frac{e\hbar}{4m^2c^2}\sigma \cdot [\nabla\Phi \times (-i\hbar\nabla - eA)]\right\}\psi_+. \tag{10}$$

The last term in the above equation is responsible for spin-orbit coupling. Since, in a static vector potential $A(r)$, the electric field is given by $E = -\nabla\Phi$, since the spin angular momentum operator is $\hat{S} = \tfrac{1}{2}\hbar\sigma$, and since the momentum operator is $\hat{p} = -i\hbar\nabla$, the spin-orbit term may be written as $-(e/2m^2c^2)\hat{S} \cdot (E \times \hat{p})$. Also, in the first line of Eq.(10), the term $-(e\hbar/2m)\sigma \cdot B = -(ge/2m)\hat{S} \cdot B$ is the Zeeman energy that couples the magnetic dipole moment of the electron $m = (ge/2m)S$ to the magnetic field $B = \nabla \times A$; here $g = 2$ is the electron's $g$-factor. The term $-i(e\hbar/4m^2c^2)\nabla\Phi \cdot (-i\hbar\nabla - eA)$, appearing on the second line of Eq.(10), has no classical counterpart and is often referred to as the Darwin term. The Darwin Hamiltonian is thus given by

$$\widehat{H}_{\text{Darwin}} = (e\hbar^2/4m^2c^2)E \cdot (\nabla - ieA/\hbar). \tag{11}$$

In first-order perturbation theory, the expectation value of the Darwin Hamiltonian for a *real-valued* wave-function $\psi_+(r)$, upon the neglect of the $E \cdot A$ term, becomes

$$\langle\psi_+|\widehat{H}_{\text{Darwin}}|\psi_+\rangle = \frac{e\hbar^2}{4m^2c^2}\iiint_{-\infty}^{\infty}\psi_+^\dagger(r)E \cdot \nabla\psi_+(r)dv = \frac{e\hbar^2}{8m^2c^2}\iiint E \cdot \nabla(\psi_+^\dagger\psi_+)dv$$

$$= -\frac{e\hbar^2}{8m^2c^2}\iiint \psi_+^\dagger\psi_+\nabla \cdot E\, dv = \frac{e\hbar^2}{8m^2c^2}\iiint \psi_+^\dagger(r)(\nabla^2\Phi)\psi_+(r)dv. \tag{12}$$

This is why the Darwin Hamiltonian is often written as $\widehat{H}_{\text{Darwin}} = (e\hbar^2/8m^2c^2)\nabla^2\Phi$, albeit without mentioning the aforementioned restrictions.

**3. Solving Dirac's equation for hydrogen-like atoms**. Let the vector potential $A(r)$ be zero, and assume a time-independent, spherically-symmetric scalar potential $\Phi(r,t) = \Phi(r)$. When the eigenstate $\psi(r,t) = \exp(-i\mathcal{E}t/\hbar)\psi(r)$ of the Dirac Hamiltonian, having a positive eigen-energy $\mathcal{E}$, is put into Eq.(5), we find

$$(\mathcal{E} - mc^2 - e\Phi)\psi_+(r) = -i\hbar c\hat{\sigma} \cdot \nabla\psi_-(r), \tag{13a}$$

$$(\mathcal{E} + mc^2 - e\Phi)\psi_-(r) = -i\hbar c\hat{\sigma} \cdot \nabla\psi_+(r). \tag{13b}$$

The operator $\hat{\boldsymbol{\sigma}} \cdot \boldsymbol{\nabla} = \begin{pmatrix} \partial_z & \partial_x - i\partial_y \\ \partial_x + i\partial_y & -\partial_z \end{pmatrix}$ may be expressed in terms of two familiar operators, $\hat{\boldsymbol{\sigma}} \cdot \boldsymbol{r}/r$ and $\hat{\boldsymbol{\sigma}} \cdot \hat{\boldsymbol{L}}$, as follows:

$$(\hat{\boldsymbol{\sigma}} \cdot \boldsymbol{r})(\hat{\boldsymbol{\sigma}} \cdot \hat{\boldsymbol{L}}) = \begin{pmatrix} z & x-iy \\ x+iy & -z \end{pmatrix} \begin{pmatrix} L_z & L_x - iL_y \\ L_x + iL_y & -L_z \end{pmatrix}$$

$$= -i\hbar \begin{pmatrix} z & x-iy \\ x+iy & -z \end{pmatrix} \begin{pmatrix} x\partial_y - y\partial_x & y\partial_z - z\partial_y - iz\partial_x + ix\partial_z \\ y\partial_z - z\partial_y + iz\partial_x - ix\partial_z & -x\partial_y + y\partial_x \end{pmatrix}$$

$$= \hbar \begin{pmatrix} z & x-iy \\ x+iy & -z \end{pmatrix}(x\partial_x + y\partial_y + z\partial_z) - \hbar(x^2+y^2+z^2)\begin{pmatrix} \partial_z & \partial_x - i\partial_y \\ \partial_x + i\partial_y & -\partial_z \end{pmatrix}$$

$$= \hbar(\hat{\boldsymbol{\sigma}} \cdot \boldsymbol{r})(\boldsymbol{r} \cdot \boldsymbol{\nabla}) - \hbar r^2 \hat{\boldsymbol{\sigma}} \cdot \boldsymbol{\nabla}$$

$$\rightarrow \quad \hat{\boldsymbol{\sigma}} \cdot \boldsymbol{\nabla} = r^{-1}(\hat{\boldsymbol{\sigma}} \cdot \boldsymbol{r}/r)(\boldsymbol{r} \cdot \boldsymbol{\nabla} - \hbar^{-1}\hat{\boldsymbol{\sigma}} \cdot \hat{\boldsymbol{L}}). \tag{14}$$

Noting that $\boldsymbol{r} \cdot \boldsymbol{\nabla} = r\partial_r$, we rewrite Eqs.(13) as

$$(\mathcal{E} - mc^2 - e\Phi)\psi_+(\boldsymbol{r}) = -icr^{-1}(\hat{\boldsymbol{\sigma}} \cdot \boldsymbol{r}/r)(\hbar r \partial_r - \hat{\boldsymbol{\sigma}} \cdot \hat{\boldsymbol{L}})\psi_-(\boldsymbol{r}), \tag{15a}$$

$$(\mathcal{E} + mc^2 - e\Phi)\psi_-(\boldsymbol{r}) = -icr^{-1}(\hat{\boldsymbol{\sigma}} \cdot \boldsymbol{r}/r)(\hbar r \partial_r - \hat{\boldsymbol{\sigma}} \cdot \hat{\boldsymbol{L}})\psi_+(\boldsymbol{r}). \tag{15b}$$

Suppose we choose the $\psi_+$ and $\psi_-$ spinors as follows:

$$\psi_+(\boldsymbol{r}) = R_+(r)\begin{bmatrix} \sqrt{(\ell+m)/(2\ell-1)}\, Y_{\ell-1,m}(\theta,\phi) \\ \sqrt{(\ell-m-1)/(2\ell-1)}\, Y_{\ell-1,m+1}(\theta,\phi) \end{bmatrix}, \tag{16a}$$

$$\psi_-(\boldsymbol{r}) = iR_-(r)\begin{bmatrix} -\sqrt{(\ell-m)/(2\ell+1)}\, Y_{\ell,m}(\theta,\phi) \\ \sqrt{(\ell+m+1)/(2\ell+1)}\, Y_{\ell,m+1}(\theta,\phi) \end{bmatrix}. \tag{16b}$$

These spinors, which share the total angular momentum quantum number $j = \ell - \tfrac{1}{2}$, are also eigenstates of $\hat{\boldsymbol{\sigma}} \cdot \hat{\boldsymbol{L}}$, albeit with different eigenvalues, namely, $(\ell-1)\hbar$ for $\psi_+$ and $-(\ell+1)\hbar$ for $\psi_-$. Subsequently, the action of $\hat{\boldsymbol{\sigma}} \cdot \boldsymbol{r}/r = \begin{bmatrix} \cos\theta & \exp(-i\phi)\sin\theta \\ \exp(i\phi)\sin\theta & -\cos\theta \end{bmatrix}$ on the angular component of $\psi_-$ turns it into that of $\psi_+$, whereas the same operator acting on the angular component of $\psi_+$ turns it into that of $\psi_-$. Recalling that $(\hat{\boldsymbol{\sigma}} \cdot \boldsymbol{r}/r)|\psi_{\ell,m+\frac{1}{2}}\rangle = -|\psi_{\ell\pm 1, m+\frac{1}{2}}\rangle$, Eqs.(15) now yield

$$rR'_-(r) + (1+\ell)R_-(r) + [\mathcal{E} - mc^2 - e\Phi(r)]rR_+(r)/(\hbar c) = 0, \tag{17a}$$

$$rR'_+(r) + (1-\ell)R_+(r) - [\mathcal{E} + mc^2 - e\Phi(r)]rR_-(r)/(\hbar c) = 0. \tag{17b}$$

Alternatively, one may exchange the angular components of $\psi_+$ and $\psi_-$ without altering the total angular momentum quantum number $j = \ell - \tfrac{1}{2}$, that is,

$$\psi_+(\boldsymbol{r}) = R_+(r)\begin{bmatrix} -\sqrt{(\ell-m)/(2\ell+1)}\, Y_{\ell,m}(\theta,\phi) \\ \sqrt{(\ell+m+1)/(2\ell+1)}\, Y_{\ell,m+1}(\theta,\phi) \end{bmatrix}, \tag{18a}$$

$$\psi_-(\mathbf{r}) = iR_-(r) \begin{bmatrix} \sqrt{(\ell+m)/(2\ell-1)}\, Y_{\ell-1,m}(\theta,\phi) \\ \sqrt{(\ell-m-1)/(2\ell-1)}\, Y_{\ell-1,m+1}(\theta,\phi) \end{bmatrix}. \tag{18b}$$

Substitution into Eqs.(15) then yields

$$rR'_-(r) + (1-\ell)R_-(r) + [\mathcal{E} - mc^2 - e\Phi(r)]rR_+(r)/(\hbar c) = 0, \tag{19a}$$

$$rR'_+(r) + (1+\ell)R_+(r) - [\mathcal{E} + mc^2 - e\Phi(r)]rR_-(r)/(\hbar c) = 0. \tag{19b}$$

To summarize, Eqs.(17) provide a pair of coupled first-order differential equations for $R_+(r)$ and $R_-(r)$ with $\ell = j + \tfrac{1}{2}$, and with the spin and orbital angular momenta of the dominant spinor $|\psi_+\rangle = |\psi_{\ell=j-\tfrac{1}{2},\, j_z=m+\tfrac{1}{2}}\rangle$ in parallel alignment. Similarly, Eqs.(19) are a pair of coupled differential equations for $R_+(r)$ and $R_-(r)$, again with $\ell = j + \tfrac{1}{2}$, but this time the spin and orbital angular momenta of the dominant spinor $|\psi_+\rangle = |\psi_{\ell=j+\tfrac{1}{2},\, j_z=m+\tfrac{1}{2}}\rangle$ are antiparallel. In both cases the orbital angular momentum quantum number $\ell$ is a positive integer ($\ell = 1, 2, 3, \ldots$).

**4. Dependence of Dirac's wave-function on the radial coordinate.** For a hydrogen-like atom, $e\Phi(r) = -Ze^2/(4\pi\varepsilon_0 r)$, where $Z$ is the number of protons in the nucleus and $\varepsilon_0$ is the permittivity of free space. Defining the parameters $\eta_1 = (mc^2 - \mathcal{E})/\hbar c$ and $\eta_2 = (mc^2 + \mathcal{E})/\hbar c$, and using the fine-structure constant $\alpha = e^2/(4\pi\varepsilon_0 \hbar c) \cong 1/137.04$, Eqs.(17) and (19) may be written as

$$R'_-(r) + (1 \pm \ell)r^{-1}R_-(r) - \eta_1 R_+(r) + Z\alpha r^{-1}R_+(r) = 0, \tag{20a}$$

$$R'_+(r) + (1 \mp \ell)r^{-1}R_+(r) - \eta_2 R_-(r) - Z\alpha r^{-1}R_-(r) = 0. \tag{20b}$$

Far from the nucleus, the terms containing $r^{-1}$ in the above equations are relatively insignificant and may be ignored, that is,

$$R'_-(r) - \eta_1 R_+(r) \cong 0; \qquad r \gg 0, \tag{21a}$$

$$R'_+(r) - \eta_2 R_-(r) \cong 0; \qquad r \gg 0. \tag{21b}$$

Substitution from one equation into the other then yields $R''_\pm(r) \cong \eta_1\eta_2 R_\pm(r)$. We thus expect both $R_+$ and $R_-$ to asymptotically approach $\exp(-\sqrt{\eta_1\eta_2}\,r)$ as $r \to \infty$. In contrast, in the vicinity of the nucleus where $r \to 0$, Eqs.(20) are approximated as

$$R'_-(r) + (1 \pm \ell)r^{-1}R_-(r) + Z\alpha r^{-1}R_+(r) \cong 0; \qquad r \to 0, \tag{22a}$$

$$R'_+(r) + (1 \mp \ell)r^{-1}R_+(r) - Z\alpha r^{-1}R_-(r) \cong 0; \qquad r \to 0. \tag{22b}$$

Substitution from one equation into the other now yields

$$r^2 R''_\pm(r) + 3rR'_\pm(r) + (1 - \ell^2 + Z^2\alpha^2)R_\pm(r) \cong 0; \qquad r \to 0. \tag{23}$$

The function $r^s$ satisfies the above equation provided that $s^2 + 2s + (1 - \ell^2 + Z^2\alpha^2) = 0$. The only acceptable value of $s$ is thus seen to be $s = \sqrt{\ell^2 - (Z\alpha)^2} - 1$. (The remaining solution would not be normalizable, except in unusual situations such as when $119 \leq Z \leq 137$ and $\ell = 1$.) We conjecture, therefore, that the general solution to Eqs.(20) must be of the following form:

$$R_+(r) = \exp(-\sqrt{\eta_1\eta_2}\,r)\, r^s \sum_{n=0}^{N} a_n r^n, \tag{24a}$$

$$R_-(r) = \exp(-\sqrt{\eta_1\eta_2}\,r)\, r^s \sum_{n=0}^{N} b_n r^n. \tag{24b}$$

Substitution into Eqs.(20) now yields

$$\sqrt{\eta_1 \eta_2} \sum_{n=0}^{N} b_n r^{s+n} - \sum_{n=0}^{N}(s+n+1\pm\ell) b_n r^{s+n-1} + \eta_1 \sum_{n=0}^{N} a_n r^{s+n} - Z\alpha \sum_{n=0}^{N} a_n r^{s+n-1} = 0, \qquad (25a)$$

$$\sqrt{\eta_1 \eta_2} \sum_{n=0}^{N} a_n r^{s+n} - \sum_{n=0}^{N}(s+n+1\mp\ell) a_n r^{s+n-1} + \eta_2 \sum_{n=0}^{N} b_n r^{s+n} + Z\alpha \sum_{n=0}^{N} b_n r^{s+n-1} = 0. \qquad (25b)$$

In these equations, the terms with the highest and lowest powers of $r$ (namely, $r^{s+N}$ and $r^{s-1}$) may be separated from the rest and treated as follows:

$$(s+1\pm\ell) b_0 + Z\alpha a_0 = 0, \qquad (26a)$$

$$(s+1\mp\ell) a_0 - Z\alpha b_0 = 0. \qquad (26b)$$

$$\sqrt{\eta_1 \eta_2}\, b_N + \eta_1 a_N = 0, \qquad (27a)$$

$$\sqrt{\eta_1 \eta_2}\, a_N + \eta_2 b_N = 0. \qquad (27b)$$

A solution for $(a_0, b_0)$ exists only if $(s+1)^2 = \ell^2 - (Z\alpha)^2$, which is the same condition as obtained previously when the asymptotic behavior of $R_{\pm}(r)$ near the origin was examined. The initial values for a recursive calculation of $(a_n, b_n)$ are thus seen to be $(a_0, b_0)$, where $a_0$ is arbitrary (needed for the eventual normalization of $R_{\pm}$), and $b_0 = [\sqrt{\ell^2 - (Z\alpha)^2} \mp \ell] a_0/(Z\alpha)$. The $\mp$ signs in this expression are respectively assigned to the wave-functions of Eqs.(16) and (18), where the upper sign ($-$) is associated with the parallel alignment of the spin and orbital angular momenta of the dominant spinor $\psi_+$, whereas the lower sign ($+$) represents their antiparallel alignment.

As for the pair $(a_N, b_N)$, Eqs.(27a) and (27b) are identical, thus yielding $a_N/b_N = -\sqrt{\eta_2/\eta_1}$. This important constraint, imposed by the necessity of terminating the polynomials of Eqs.(24) at $n = N$, plays a crucial role in quantizing the energy $\mathcal{E}$ of the electron, as will be explained below.

Having removed the first and last terms of the polynomials from Eqs.(25), we now rearrange these equations in order to arrive at a recursion relation for the remaining pairs of coefficients $(a_n, b_n)$. The rearranged equations are

$$\sqrt{\eta_1 \eta_2} \sum_{n=0}^{N-1} b_n r^{s+n} - \sum_{n=0}^{N-1}(s+n+2\pm\ell) b_{n+1} r^{s+n} + \eta_1 \sum_{n=0}^{N-1} a_n r^{s+n} - Z\alpha \sum_{n=0}^{N-1} a_{n+1} r^{s+n} = 0, \qquad (28a)$$

$$\sqrt{\eta_1 \eta_2} \sum_{n=0}^{N-1} a_n r^{s+n} - \sum_{n=0}^{N-1}(s+n+2\mp\ell) a_{n+1} r^{s+n} + \eta_2 \sum_{n=0}^{N-1} b_n r^{s+n} + Z\alpha \sum_{n=0}^{N-1} b_{n+1} r^{s+n} = 0. \qquad (28b)$$

Setting to zero the coefficients of the various powers of $r$ from $s$ to $(s+N-1)$ now yields

$$\begin{pmatrix} \eta_1 & \sqrt{\eta_1 \eta_2} \\ \sqrt{\eta_1 \eta_2} & \eta_2 \end{pmatrix} \begin{pmatrix} a_n \\ b_n \end{pmatrix} = \begin{pmatrix} Z\alpha & s+n+2\pm\ell \\ s+n+2\mp\ell & -Z\alpha \end{pmatrix} \begin{pmatrix} a_{n+1} \\ b_{n+1} \end{pmatrix}; \quad (0 \le n \le N-1). \qquad (29)$$

Noting that the determinant of the $2 \times 2$ matrix appearing on the left-hand-side of Eq.(29) is zero, it is clear that the right-hand-side of the equation must be a two-element vector whose elements are in the ratio of $\sqrt{\eta_1/\eta_2}$. In particular, for $n = N-1$, considering that $a_N/b_N$ is already determined to be $-\sqrt{\eta_2/\eta_1}$, we must have

$$\frac{Z\alpha - (s+N+1\pm\ell)\sqrt{\eta_1/\eta_2}}{(s+N+1\mp\ell) + Z\alpha\sqrt{\eta_1/\eta_2}} = \sqrt{\eta_1/\eta_2}. \tag{30}$$

Recalling that $\eta_1 = (mc^2 - \mathcal{E})/\hbar c$, $\eta_2 = (mc^2 + \mathcal{E})/\hbar c$, $s + 1 = \sqrt{\ell^2 - (Z\alpha)^2}$, and $\ell = j + \tfrac{1}{2}$, Eq.(30) yields

$$\mathcal{E}_N = \left[1 + \left(\frac{Z\alpha}{N + \sqrt{(j+\tfrac{1}{2})^2 - (Z\alpha)^2}}\right)^2\right]^{-\tfrac{1}{2}} mc^2. \tag{31}$$

The subscript of $\mathcal{E}_N$ in the above equation serves as a reminder that this energy is associated with polynomials of degree $N$ in the expressions of $R_\pm(r)$ given in Eqs.(24). Note that $\mathcal{E}_N$, while positive, is less than $mc^2$ for all values of $Z$, $N$, and $j$. Recall that $j$, the total angular momentum quantum number, is a positive half-integer greater than or equal to $\tfrac{1}{2}$. (If the value of $Z$ happens to exceed 137, then the minimum value of $j$ will jump to $3/2$.) Had we chosen to put a minus sign before the expression of $\mathcal{E}_N$ in Eq.(31), it would have corresponded to negative-energy solutions of Dirac's equation. For positive $\mathcal{E}_N$ values we have $\eta_2 \gg \eta_1$, which results in the magnitude of $a_N$ being substantially greater than that of $b_N$; the opposite, of course, is true for negative $\mathcal{E}_N$ values.

The principal quantum number in Schrödinger's model of hydrogen-like atoms can now be shown to relate to Dirac's $N$ and $j$ via $n_{\text{schr}} = N + j + \tfrac{1}{2}$. This may be readily seen by approximating Eq.(31) as follows:

$$mc^2 - \mathcal{E}_N \cong mc^2 - \left[1 - \tfrac{1}{2}\left(\frac{Z\alpha}{N+\sqrt{(j+\tfrac{1}{2})^2-(Z\alpha)^2}}\right)^2\right] mc^2 \cong \frac{mc^2 Z^2 \alpha^2}{2(N+j+\tfrac{1}{2})^2}. \tag{32}$$

Substituting in the above equation the numerical values of $Z = 1$, $mc^2 \cong 0.511 \times 10^6$ eV, and $\alpha \cong 1/137$ (corresponding to atomic hydrogen) yields: $mc^2 - \mathcal{E}_N \cong 13.6/(N + j + \tfrac{1}{2})^2$ eV.

The case of $N = 0$ is particularly interesting because, in Eqs.(26) and (27), we have two constraints on the same pair of coefficients, $(a_0, b_0)$. According to Eqs.(26),

$$a_0/b_0 = -\left[\sqrt{\ell^2 - (Z\alpha)^2} \pm \ell\right]/Z\alpha, \tag{33}$$

where the upper sign (+) would make $|a_0| \gg |b_0|$, whereas the lower sign (−) would result in $|a_0| \ll |b_0|$. Similarly, Eqs.(27) require that $a_0/b_0 = -\sqrt{\eta_2/\eta_1}$, where the positive-energy solution (i.e., $\mathcal{E}_0 > 0$) would yield $|a_0| \gg |b_0|$, while the negative-energy solution would result in $|a_0| \ll |b_0|$. Consistency demands that these two situations be identical. Indeed, setting $N = 0$ in Eq.(31) and evaluating $\eta_2/\eta_1$ yields

$$a_0/b_0 = -\sqrt{\eta_2/\eta_1} = -\left[\ell \pm \sqrt{\ell^2 - (Z\alpha)^2}\right]/Z\alpha. \tag{34}$$

There is agreement between Eqs.(33) and (34) when the upper sign (+) applies, but there is disagreement in case of the lower sign (−). It is thus seen that, when $N = 0$, the only acceptable wave-function has positive-energy (i.e., $\mathcal{E}_0 > 0$) and is given by Eqs.(16), where the orbital angular momentum quantum number of $\psi_+(\mathbf{r})$ is $\ell - 1 = j - \tfrac{1}{2}$. This is consistent with Schrödinger's model, in which $n_{\text{schr}} = j + \tfrac{1}{2} = \ell_{\text{schr}} + 1$. In other words, when $N = 0$, all half-integer values of $j \geq \tfrac{1}{2}$ are allowed, but anti-parallel alignment of spin and orbital angular momenta is prohibited.[†]

The stage is now set for invoking the recursion relation, Eq.(29), to calculate $(a_n, b_n)$ from the preceding coefficients $(a_{n-1}, b_{n-1})$. Inverting the $2 \times 2$ matrix on the right-hand-side of Eq.(29) and using the identity $s + 1 = \sqrt{\ell^2 - (Z\alpha)^2}$, we arrive at

---

[†] The same conclusions can be reached more directly by solving Eqs.(25) for $(a_0, b_0)$ in the special case of $N = 0$.

$$\binom{a_n}{b_n} = \frac{\sqrt{\eta_1}\,a_{n-1} + \sqrt{\eta_2}\,b_{n-1}}{n^2 + 2n\sqrt{\ell^2 - (Z\alpha)^2}} \begin{pmatrix} Z\alpha & n \pm \ell + \sqrt{\ell^2 - (Z\alpha)^2} \\ n \mp \ell + \sqrt{\ell^2 - (Z\alpha)^2} & -Z\alpha \end{pmatrix} \binom{\sqrt{\eta_1}}{\sqrt{\eta_2}}; \quad (1 \le n \le N). \quad (35)$$

Considering that $\sqrt{\eta_2} \gg \sqrt{\eta_1}$ for $\mathcal{E}_N > 0$, it is seen that the magnitude of each $a_n$ is substantially greater than that of $b_n$, irrespective of the initial choice of $(a_o, b_o) = a_o[1, (\sqrt{\ell^2 - (Z\alpha)^2} \mp \ell)/(Z\alpha)]$. This is also true of $(a_o, b_o)$ itself, provided that the upper sign $(-)$ is chosen for $b_o$, although the choice of the lower sign $(+)$ will result in $|b_o| \gg |a_o|$. Apparently, in cases where the spin and orbital angular momenta of the dominant spinor $\psi_+(r)$ are anti-parallel, the lowest-power term $b_o r^s$ of $\psi_-(r)$ far exceeds the corresponding term $a_o r^s$ of $\psi_+(r)$. The effects of this term, however, are significant only in the immediate vicinity of the nucleus of the atom.

**5. Some useful identities**. The following identities are found to be useful in various calculations.

$$\frac{\mathcal{E}_N}{mc^2} = \frac{N + \sqrt{\ell^2 - (Z\alpha)^2}}{[N^2 + \ell^2 + 2N\sqrt{\ell^2 - (Z\alpha)^2}]^{1/2}}. \quad (36)$$

$$\sqrt{\eta_1 \eta_2} = \frac{(mc^2 Z\alpha)/\hbar c}{[N^2 + \ell^2 + 2N\sqrt{\ell^2 - (Z\alpha)^2}]^{1/2}}. \quad (37)$$

$$\frac{\eta_2 - \eta_1}{2\sqrt{\eta_1 \eta_2}} = \frac{\mathcal{E}_N/mc^2}{\sqrt{1 - (\mathcal{E}_N/mc^2)^2}} = \frac{N + \sqrt{\ell^2 - (Z\alpha)^2}}{Z\alpha}. \quad (38)$$

$$\left(\sqrt{\eta_1} \pm \sqrt{\eta_2}\right)^2 = \eta_1 + \eta_2 \pm 2\sqrt{\eta_1 \eta_2} = 2(mc^2/\hbar c) \pm 2\sqrt{\eta_1 \eta_2}. \quad (39)$$

$$Z\alpha \sqrt{\eta_1/\eta_2} = [N^2 + \ell^2 + 2N\sqrt{\ell^2 - (Z\alpha)^2}]^{1/2} - [N + \sqrt{\ell^2 - (Z\alpha)^2}]. \quad (40)$$

$$Z\alpha \sqrt{\eta_2/\eta_1} = [N^2 + \ell^2 + 2N\sqrt{\ell^2 - (Z\alpha)^2}]^{1/2} + [N + \sqrt{\ell^2 - (Z\alpha)^2}]. \quad (41)$$

**6. Explicit formula for the $(a_n, b_n)$ coefficients**. Let us rearrange Eq.(35) in the following way:

$$\binom{a_n}{b_n} = \frac{\left[\frac{1}{Z\alpha}\binom{\sqrt{\ell^2-(Z\alpha)^2} \pm \ell}{-Z\alpha}\right]\left(-\sqrt{\ell^2-(Z\alpha)^2} \pm \ell \quad Z\alpha\right) + n\binom{0\;1}{1\;0}\binom{\sqrt{\eta_1}}{\sqrt{\eta_2}}}{n[2\sqrt{\ell^2-(Z\alpha)^2}+n]} \times (\sqrt{\eta_1}\;\;\sqrt{\eta_2})\binom{a_{n-1}}{b_{n-1}}. \quad (42)$$

The above matrix equation is further simplified upon reducing the product of a row-vector and a column-vector to a scalar, namely,

$$\binom{a_n}{b_n} = \frac{\left(\sqrt{\eta_2} + [\pm(\ell/Z\alpha) - \sqrt{(\ell/Z\alpha)^2 - 1}]\sqrt{\eta_1}\right)\binom{\sqrt{\ell^2-(Z\alpha)^2} \pm \ell}{-Z\alpha} + n\binom{\sqrt{\eta_2}}{\sqrt{\eta_1}}}{n[2\sqrt{\ell^2-(Z\alpha)^2}+n]} \times (\sqrt{\eta_1}\;\;\sqrt{\eta_2})\binom{a_{n-1}}{b_{n-1}}. \quad (43)$$

The same recursion relation applied to the last term on the right-hand-side of Eq.(43) yields

$$(\sqrt{\eta_1}\;\;\sqrt{\eta_2})\binom{a_{n-1}}{b_{n-1}} = \frac{(\eta_1 - \eta_2)(Z\alpha) + 2\sqrt{\eta_1\eta_2}\,[\sqrt{\ell^2-(Z\alpha)^2}+(n-1)]}{(n-1)[2\sqrt{\ell^2-(Z\alpha)^2}+(n-1)]} \times (\sqrt{\eta_1}\;\;\sqrt{\eta_2})\binom{a_{n-2}}{b_{n-2}}. \quad (44)$$

It is now easy to show, with the aid of Eq.(31), that $(\eta_1 - \eta_2)Z\alpha = -2\sqrt{\eta_1 \eta_2}\,[N + \sqrt{\ell^2 - (Z\alpha)^2}]$. Upon substitution into Eq.(44), we find

$$(\sqrt{\eta_1} \quad \sqrt{\eta_2})\begin{pmatrix}a_{n-1}\\b_{n-1}\end{pmatrix} = -\frac{2\sqrt{\eta_1\eta_2}\,[N-(n-1)]}{(n-1)[2\sqrt{\ell^2-(Z\alpha)^2}+(n-1)]} \times (\sqrt{\eta_1} \quad \sqrt{\eta_2})\begin{pmatrix}a_{n-2}\\b_{n-2}\end{pmatrix}. \quad (45)$$

Repeating the above recursion all the way down to $(a_0, b_0)$ now yields

$$(\sqrt{\eta_1} \quad \sqrt{\eta_2})\begin{pmatrix}a_{n-1}\\b_{n-1}\end{pmatrix} = \frac{(-2\sqrt{\eta_1\eta_2})^{n-1}(N-1)!/(N-n)!}{(n-1)![2\sqrt{\ell^2-(Z\alpha)^2}+1]\cdots[2\sqrt{\ell^2-(Z\alpha)^2}+(n-1)]} \times (\sqrt{\eta_1} \quad \sqrt{\eta_2})\begin{pmatrix}a_0\\b_0\end{pmatrix}. \quad (46)$$

Substitution for the initial coefficients $(a_0, b_0) = a_0\left(1, \left[\sqrt{\ell^2 - (Z\alpha)^2} \mp \ell\right]/Z\alpha\right)$ followed by insertion of Eq.(46) into Eq.(43) finally yields

$$\begin{pmatrix}a_n\\b_n\end{pmatrix} = \frac{a_0}{2N}\binom{N}{n}\frac{(-2\sqrt{\eta_1\eta_2})^n}{[2\sqrt{\ell^2-(Z\alpha)^2}+1]\cdots[2\sqrt{\ell^2-(Z\alpha)^2}+n]}\begin{pmatrix}(2N-n)-n\sqrt{\eta_2/\eta_1}\,[\sqrt{\ell^2-(Z\alpha)^2}\mp\ell]/Z\alpha\\-n\sqrt{\eta_1/\eta_2}+(2N-n)[\sqrt{\ell^2-(Z\alpha)^2}\mp\ell]/Z\alpha\end{pmatrix}. \quad (47)$$

In these equations, $\ell = j + \tfrac{1}{2}$, and the upper and lower signs correspond, respectively, to the parallel and antiparallel alignment of the spin and orbital angular momenta of the dominant spinor $\psi_+(\mathbf{r})$. As a check of validity of the final result, note that the ratio $a_N/b_N$ indeed equals $-\sqrt{\eta_2/\eta_1}$. The term appearing in the denominator in Eq.(47) may be expressed as

$$[2\sqrt{\ell^2-(Z\alpha)^2}+1]\cdots[2\sqrt{\ell^2-(Z\alpha)^2}+n] = \Gamma(2\sqrt{\ell^2-(Z\alpha)^2}+n+1)/\Gamma(2\sqrt{\ell^2-(Z\alpha)^2}+1). \quad (48)$$

In this way, Eq.(47) becomes formally valid for $n = 0$ as well.

**7. Probability density, probability current, and normalization of Dirac's spinor**. Starting with Dirac's equation, Eq.(4), let us write down both the equation and its complex conjugate, then multiply (on the left) the first equation with $\psi^\dagger(\mathbf{r},t) = \psi^{*T}(\mathbf{r},t)$ and the second one with $\psi^T(\mathbf{r},t)$. We will have

$$\psi^\dagger(i\hbar\partial_t - e\Phi)\psi = c\psi^\dagger\boldsymbol{\alpha}\cdot(-i\hbar\boldsymbol{\nabla} - e\mathbf{A})\psi + mc^2\psi^\dagger\beta\psi, \quad (49a)$$

$$\psi^T(-i\hbar\partial_t - e\Phi)\psi^* = c\psi^T\boldsymbol{\alpha}^*\cdot(+i\hbar\boldsymbol{\nabla} - e\mathbf{A})\psi^* + mc^2\psi^T\beta^*\psi^*. \quad (49b)$$

Note that, since the matrix $\beta$ is real-valued, $\psi^\dagger\beta\psi = \psi^T\beta^*\psi^* = |\psi_1|^2 + |\psi_2|^2 - |\psi_3|^2 - |\psi_4|^2$. Similarly, $\psi^\dagger\Phi\psi = \psi^T\Phi\psi^* = \Phi(\mathbf{r},t)(|\psi_1|^2+|\psi_2|^2+|\psi_3|^2+|\psi_4|^2)$. Also, considering that $\alpha_x$, $\alpha_y$, and $\alpha_z$ are Hermitian matrices (that is, $\alpha_x^\dagger = \alpha_x$, etc.), we may write

$$\psi^\dagger(A_x\alpha_x + A_y\alpha_y + A_z\alpha_z)\psi = \psi^T(A_x\alpha_x + A_y\alpha_y^* + A_z\alpha_z)\psi^*$$

$$= A_x(\psi_1\psi_4^* + \psi_2\psi_3^* + \psi_3\psi_2^* + \psi_4\psi_1^*)$$

$$+ iA_y(\psi_1\psi_4^* - \psi_2\psi_3^* + \psi_3\psi_2^* - \psi_4\psi_1^*)$$

$$+ A_z(\psi_1\psi_3^* - \psi_2\psi_4^* + \psi_3\psi_1^* - \psi_4\psi_2^*). \quad (50)$$

Consequently, subtracting Eq.(49b) from Eq.(49a) yields

$$i\hbar(\psi^\dagger\partial_t\psi + \psi^T\partial_t\psi^*) = -i\hbar c\psi^\dagger(\alpha_x\partial_x + \alpha_y\partial_y + \alpha_z\partial_z)\psi - i\hbar c\psi^T(\alpha_x\partial_x + \alpha_y^*\partial_y + \alpha_z\partial_z)\psi^*$$

$$= -i\hbar c[(\psi^\dagger\alpha_x\partial_x\psi + \psi^T\alpha_x\partial_x\psi^*) + (\psi^\dagger\alpha_y\partial_y\psi + \psi^T\alpha_y^*\partial_y\psi^*) + (\psi^\dagger\alpha_z\partial_z\psi + \psi^T\alpha_z\partial_z\psi^*)]. \quad (51)$$

Noting that $\psi^\dagger\psi = |\psi_1|^2 + |\psi_2|^2 + |\psi_3|^2 + |\psi_4|^2 = \psi^T\psi^*$, the left-hand-side of Eq.(51) is readily seen to be equal to $i\hbar\partial_t(\psi^\dagger\psi)$. The probability-density of finding the electron at $(r,t)$ may thus be denoted by $\mathcal{P}(r,t) = \psi^\dagger\psi$. Appearing implicitly on the right-hand-side of Eq.(51) are the components of the probability current-density $\boldsymbol{J} = \mathcal{J}_x\hat{x} + \mathcal{J}_y\hat{y} + \mathcal{J}_z\hat{z}$, which are defined as follows:

$$\mathcal{J}_x(r,t) = c\psi^\dagger\alpha_x\psi = c\psi^T\alpha_x\psi^* = c(\psi_1\psi_4^* + \psi_2\psi_3^* + \psi_3\psi_2^* + \psi_4\psi_1^*), \tag{52a}$$

$$\mathcal{J}_y(r,t) = c\psi^\dagger\alpha_y\psi = c\psi^T\alpha_y^*\psi^* = ic(\psi_1\psi_4^* - \psi_2\psi_3^* + \psi_3\psi_2^* - \psi_4\psi_1^*), \tag{52b}$$

$$\mathcal{J}_z(r,t) = c\psi^\dagger\alpha_z\psi = c\psi^T\alpha_z\psi^* = c(\psi_1\psi_3^* - \psi_2\psi_4^* + \psi_3\psi_1^* - \psi_4\psi_2^*). \tag{52c}$$

The continuity equation $\boldsymbol{\nabla}\cdot\boldsymbol{J} + \partial_t\mathcal{P} = 0$ relating probability current and probability density thus emerges directly from Eq.(51). The integral over the entire space of the probability-density function, that is, $\iiint_{-\infty}^{\infty}\mathcal{P}(r,t)\,dxdydz$, which must be equal to unity at all times, is readily seen to be time-independent.

The normalization condition may now be applied to the energy eigen-functions of hydrogen-like atoms to determine the remaining coefficient $a_0$. For the wave-function of Eq.(16) we have

$$\iiint_{-\infty}^{\infty}(|\psi_{+\uparrow}|^2 + |\psi_{+\downarrow}|^2 + |\psi_{-\uparrow}|^2 + |\psi_{-\downarrow}|^2)\,dxdydz$$

$$= \int_{r=0}^{\infty}\int_{\theta=0}^{\pi}\int_{\phi=0}^{2\pi}|R_+(r)|^2\left[|\alpha_{+\uparrow}Y_{\ell-1,m}(\theta,\phi)|^2 + |\alpha_{+\downarrow}Y_{\ell-1,m+1}(\theta,\phi)|^2\right]r^2\sin\theta\,drd\theta d\phi$$

$$+ \int_{r=0}^{\infty}\int_{\theta=0}^{\pi}\int_{\phi=0}^{2\pi}|R_-(r)|^2\left[|\alpha_{-\uparrow}Y_{\ell,m}(\theta,\phi)|^2 + |\alpha_{-\downarrow}Y_{\ell,m+1}(\theta,\phi)|^2\right]r^2\sin\theta\,drd\theta d\phi$$

$$= \int_0^\infty r^2[|R_+(r)|^2 + |R_-(r)|^2]\,dr$$

$$= \int_0^\infty \exp(-2\sqrt{\eta_1\eta_2}\,r)\,r^{2(s+1)}[(\textstyle\sum_{n=0}^N a_n r^n)^2 + (\textstyle\sum_{n=0}^N b_n r^n)^2]\,dr$$

$$= \int_0^\infty \exp(-2\sqrt{\eta_1\eta_2}\,r)\,r^{2\sqrt{\ell^2-(Z\alpha)^2}}\sum_{k=0}^{2N}\left[\sum_{n=\max(0,k-N)}^{\min(N,k)}(a_n a_{k-n} + b_n b_{k-n})\right]r^k\,dr$$

$$= \sum_{k=0}^{2N}\left[\sum_{n=\max(0,k-N)}^{\min(N,k)}(a_n a_{k-n} + b_n b_{k-n})\right]\Gamma(2\sqrt{\ell^2-(Z\alpha)^2} + k + 1)/(2\sqrt{\eta_1\eta_2})^{2\sqrt{\ell^2-(Z\alpha)^2}+k+1}$$

<div style="text-align:right">Gradshteyn & Ryzhik[6] **3.381**-4</div>

$$= 1. \tag{53}$$

Needless to say, the above equation applies also to the wave-function of Eq.(18). The coefficients $(a_n, b_n)$ are given by Eq.(47).

**8. Examples**. As a first example, let us consider the state $2p_{3/2}$ of the hydrogen atom ($Z = 1$), where $j = 3/2$, $\ell = 2$, $N = 0$, and where the spin and orbital angular momenta are parallel to each other. From Eq.(36), we find

$$\mathcal{E}/(mc^2) = \sqrt{1 - (\alpha/2)^2} = 1 - \tfrac{\alpha^2}{8}\left(1 + \tfrac{\alpha^2}{16} + \tfrac{\alpha^4}{128} + \cdots\right). \tag{54}$$

The normalization condition, Eq.(53), yields the remaining unknown coefficient, $a_0$, as follows:

$$(a_o^2 + b_o^2)\Gamma(2\sqrt{\ell^2 - \alpha^2} + 1)/(2\sqrt{\eta_1\eta_2})^{2\sqrt{\ell^2-\alpha^2}+1} = 1 \quad \rightarrow \quad a_o^2 = \frac{\alpha^2(2\sqrt{\eta_1\eta_2})^{2\sqrt{\ell^2-\alpha^2}+1}}{2\ell(\ell - \sqrt{\ell^2-\alpha^2})\Gamma(2\sqrt{\ell^2-\alpha^2}+1)}. \quad (55)$$

The dependence of the wave-function on the radial coordinates is thus given by Eqs.(24), with $s = 2\sqrt{1 - (\alpha/2)^2} - 1$ and $b_o = (2/\alpha)[\sqrt{1 - (\alpha/2)^2} - 1]a_o$. The dominant spinor is that of Eq.(16a), with the allowed $m$ values being $+1, 0, -1, -2$, corresponding to $j_z = \pm 3/2$ and $\pm 1/2$.

As a second example, consider the state $2p_{1/2}$ of the hydrogen atom, where $j = \frac{1}{2}$, $\ell = 1$, and $N = 1$, with the spin aligned anti-parallel to the orbital angular momentum. From Eq.(36) we find

$$\mathcal{E}/(mc^2) = [(1 + \sqrt{1 - \alpha^2})/2]^{1/2} = 1 - \frac{\alpha^2}{8}\left(1 + \frac{5\alpha^2}{16} + \frac{21\alpha^4}{128} + \cdots\right). \quad (56)$$

The normalization condition, Eq.(53), in conjunction with Eqs.(47) and (48), now yields

$$(a_o^2 + b_o^2)\Gamma(2\sqrt{\ell^2 - \alpha^2} + 1)/(2\sqrt{\eta_1\eta_2})^{2\sqrt{\ell^2-\alpha^2}+1} + 2(a_o a_1 + b_o b_1)\Gamma(2\sqrt{\ell^2 - \alpha^2} + 2)/(2\sqrt{\eta_1\eta_2})^{2\sqrt{\ell^2-\alpha^2}+2}$$
$$+ (a_1^2 + b_1^2)\Gamma(2\sqrt{\ell^2 - \alpha^2} + 3)/(2\sqrt{\eta_1\eta_2})^{2\sqrt{\ell^2-\alpha^2}+3} = 1. \quad (57)$$

where

$$a_o^2 + b_o^2 = [2\ell(\ell + \sqrt{\ell^2 - \alpha^2})/\alpha^2]a_o^2, \quad (57a)$$

$$a_o a_1 + b_o b_1 = \frac{[(\eta_1 + \eta_2) - 2(\ell/\alpha)\sqrt{\eta_1\eta_2}](\ell + \sqrt{\ell^2-\alpha^2})}{\alpha(2\sqrt{\ell^2-\alpha^2} + 1)} a_o^2, \quad (57b)$$

$$a_1^2 + b_1^2 = \frac{(\eta_1 + \eta_2)\{(\eta_1 - \eta_2) - 2[\sqrt{\eta_1\eta_2} - (\ell/\alpha)\eta_2](\ell + \sqrt{\ell^2-\alpha^2})/\alpha\}}{(2\sqrt{\ell^2-\alpha^2} + 1)^2} a_o^2. \quad (57c)$$

The coefficient $a_o$ can thus be computed from Eqs.(57), then substituted into Eqs.(47) and (48), along with $s = \sqrt{1 - \alpha^2} - 1$, to compute the remaining coefficients $b_o$, $a_1$, and $b_1$. Subsequently, Eqs.(24) specify the radial dependence $R_\pm(r)$ of $\psi_\pm(r)$, and Eqs.(18) yield the complete Dirac wave-function for the electron in the $2p_{1/2}$ state of the hydrogen atom. The allowed $m$ values in this case are 0 and $-1$, corresponding to $j_z = \pm \frac{1}{2}$.

**9. Concluding remarks**. Beginning with Dirac's relativistic equation, we have shown in this paper that the complete and rigorous solution of the equation for hydrogen (and hydrogen-like) atoms is all that is needed to understand, among other things, the existence and strength of the coupling between the spin and orbital angular momenta of the electron. While the existence of exact solutions for the electron's wave-function enables us to compute its various characteristics (e.g., the average radius of the electron's orbit in its manifold energy eigen-states, such as the $2p_{3/2}$ and $2p_{1/2}$ states discussed in Sec.8), it is probably a futile exercise to try to relate these features to the Bohr model of the hydrogen atom in an attempt to extract useful quantitative information about the spin-orbit coupling energy.

The interaction between an electron's magnetic dipole moment (rooted in its spin) and the magnetic field experienced by the electron in its rest-frame (due to the electric charge of the nucleus) is universally acknowledged as the root cause of the spin-orbit coupling energy of the electron. Agreement between theory and experiment, however, has required accounting for a missing ½ factor in the theoretical formula derived on this basis. Traditionally, this missing factor has been associated with a relativistic effect known as the Thomas precession.[1] As we explained in a previous publication,[2] there are reasons to question the validity of pinning the blame for the missing ½ factor onto the Thomas precession mechanism. Instead, we suggested that, within the Bohr model of the hydrogen atom, the balance of the electron's kinetic and potential energies could account for the missing factor provided that, during a spin-flip transition, the orbital radius of the electron remains unchanged.[2]

While the exact solution of Dirac's equation provides the opportunity to test the above hypothesis for the average (i.e., expected value) of the orbital radius of the electron—say, by comparing the average radii of the $2p_{3/2}$ and $2p_{1/2}$ states—we believe that neither the affirmation, nor the refutation, of the aforementioned hypothesis will substantively strengthen our understanding of the spin-orbit interaction. It is our contention that the Dirac equation is simple enough, and its solution for the hydrogen atom rich, elegant, and powerful enough, that any attempts at a deeper understanding of the magnetic moment of the electron in its interaction with the electric field of the nucleus should begin with Dirac's equation, rather than with the ad hoc, semi-classical, and far less sophisticated—albeit historically significant—Bohr's model of the hydrogen atom.

# Appendix
# Angular Momentum in Quantum Mechanics

**A1. Angular momentum operators**. The classical definition of angular momentum $\boldsymbol{L} = \boldsymbol{r} \times \boldsymbol{p}$ leads to the quantum-mechanical operators $\hat{L}_x, \hat{L}_y, \hat{L}_z, \hat{\boldsymbol{L}}$ and $\hat{L}^2$, as follows:

$$\hat{L}_x = -i\hbar(y\partial_z - z\partial_y), \qquad \hat{L}_y = -i\hbar(z\partial_x - x\partial_z), \qquad \hat{L}_z = -i\hbar(x\partial_y - y\partial_x),$$

$$\hat{\boldsymbol{L}} = \hat{L}_x \hat{\boldsymbol{x}} + \hat{L}_y \hat{\boldsymbol{y}} + \hat{L}_z \hat{\boldsymbol{z}}, \qquad \hat{L}^2 = \hat{L}_x^2 + \hat{L}_y^2 + \hat{L}_z^2. \tag{A1}$$

Note in Eq.(A1) that $\hat{\boldsymbol{x}}, \hat{\boldsymbol{y}}, \hat{\boldsymbol{z}}$ are *not* operators but rather unit-vectors along the Cartesian coordinate axes. The various commutation relations involving angular momentum operators are listed below.

$$[\hat{L}_x, \hat{L}_y] = \hat{L}_x \hat{L}_y - \hat{L}_y \hat{L}_x = -\hbar^2[(y\partial_z - z\partial_y)(z\partial_x - x\partial_z) - (z\partial_x - x\partial_z)(y\partial_z - z\partial_y)]$$

$$= -\hbar^2[(y\partial_x + yz\partial_{zx}^2 - yx\partial_z^2 - z^2\partial_{yx}^2 + zx\partial_{yz}^2) - (zy\partial_{xz}^2 - z^2\partial_{xy}^2 - xy\partial_z^2 + x\partial_y + xz\partial_{zy}^2)]$$

$$= -\hbar^2(y\partial_x - x\partial_y) = i\hbar \hat{L}_z. \tag{A2a}$$

$$[\hat{L}_y, \hat{L}_z] = i\hbar \hat{L}_x. \tag{A2b}$$

$$[\hat{L}_z, \hat{L}_x] = i\hbar \hat{L}_y. \tag{A2c}$$

$$[\hat{L}^2, \hat{L}_x] = [\hat{L}_x^2, \hat{L}_x] + [\hat{L}_y^2, \hat{L}_x] + [\hat{L}_z^2, \hat{L}_x] = (\hat{L}_y \hat{L}_y \hat{L}_x - \hat{L}_x \hat{L}_y^2) + (\hat{L}_z \hat{L}_z \hat{L}_x - \hat{L}_x \hat{L}_z^2)$$

$$= [\hat{L}_y(\hat{L}_x \hat{L}_y - i\hbar \hat{L}_z) - \hat{L}_x \hat{L}_y^2] + [\hat{L}_z(\hat{L}_x \hat{L}_z + i\hbar \hat{L}_y) - \hat{L}_x \hat{L}_z^2]$$

$$= [(\hat{L}_x \hat{L}_y - i\hbar \hat{L}_z)\hat{L}_y - i\hbar \hat{L}_y \hat{L}_z - \hat{L}_x \hat{L}_y^2] + [(\hat{L}_x \hat{L}_z + i\hbar \hat{L}_y)\hat{L}_z + i\hbar \hat{L}_z \hat{L}_y - \hat{L}_x \hat{L}_z^2] = 0.$$

$$\tag{A2d}$$

$$[\hat{L}^2, \hat{L}_y] = 0. \tag{A2e}$$

$$[\hat{L}^2, \hat{L}_z] = 0. \tag{A2f}$$

$$[\hat{L}_x, \hat{p}_x] = -\hbar^2[(y\partial_z - z\partial_y)\partial_x - \partial_x(y\partial_z - z\partial_y)] = 0. \tag{A2g}$$

$$[\hat{L}_x, \hat{p}_y] = -\hbar^2[(y\partial_z - z\partial_y)\partial_y - \partial_y(y\partial_z - z\partial_y)]$$

$$= -\hbar^2(y\partial_{zy}^2 - z\partial_y^2 - \partial_z - y\partial_{yz}^2 + z\partial_y^2) = i\hbar \hat{p}_z. \tag{A2h}$$

$$[\hat{L}_x, \hat{p}_z] = -i\hbar \hat{p}_y. \tag{A2i}$$

$$[\hat{L}_x, \hat{p}^2] = [\hat{L}_x, \hat{p}_x^2] + [\hat{L}_x, \hat{p}_y^2] + [\hat{L}_x, \hat{p}_z^2]$$

$$= (\hat{L}_x \hat{p}_x \hat{p}_x - \hat{p}_x^2 \hat{L}_x) + (\hat{L}_x \hat{p}_y \hat{p}_y - \hat{p}_y^2 \hat{L}_x) + (\hat{L}_x \hat{p}_z \hat{p}_z - \hat{p}_z^2 \hat{L}_x)$$

$$= (\hat{p}_x \hat{L}_x \hat{p}_x - \hat{p}_x^2 \hat{L}_x) + [(\hat{p}_y \hat{L}_x + i\hbar \hat{p}_z)\hat{p}_y - \hat{p}_y^2 \hat{L}_x] + [(\hat{p}_z \hat{L}_x - i\hbar \hat{p}_y)\hat{p}_z - \hat{p}_z^2 \hat{L}_x]$$

$$= (\hat{p}_x^2 \hat{L}_x - \hat{p}_x^2 \hat{L}_x) + [\hat{p}_y(\hat{p}_y \hat{L}_x + i\hbar \hat{p}_z) + i\hbar \hat{p}_z \hat{p}_y - \hat{p}_y^2 \hat{L}_x]$$

$$+ [\hat{p}_z(\hat{p}_z \hat{L}_x - i\hbar \hat{p}_y) - i\hbar \hat{p}_y \hat{p}_z - \hat{p}_z^2 \hat{L}_x] = 0. \tag{A2j}$$

$$[\hat{L}_y, \hat{p}^2] = 0, \qquad [\hat{L}_z, \hat{p}^2] = 0, \qquad [\hat{L}^2, \hat{p}^2] = [\hat{L}_x^2, \hat{p}^2] + [\hat{L}_y^2, \hat{p}^2] + [\hat{L}_z^2, \hat{p}^2] = 0. \tag{A2k}$$

In the case of the spherically symmetric scalar potential $\Phi(r)$, where $r = \sqrt{x^2 + y^2 + z^2}$ denotes the radial distance from the origin, the commutation relation between $\Phi(r)$ and the angular momentum operators will be

$$[\hat{L}_x, \Phi(r)] = -i\hbar[(y\partial_z - z\partial_y)\Phi(r) - \Phi(r)(y\partial_z - z\partial_y)]$$

$$= -i\hbar\{y[\partial_z\Phi(r)] + y\Phi(r)\partial_z - z[\partial_y\Phi(r)] - z\Phi(r)\partial_y - y\Phi(r)\partial_z + z\Phi(r)\partial_y\}$$

$$= -i\hbar\{y\partial_r[\Phi(r)](\partial_z r) - z\partial_r[\Phi(r)](\partial_y r)\} = -i\hbar\{y\partial_r[\Phi(r)](z/r) - z\partial_r[\Phi(r)](y/r)\} = 0,$$

$$[\hat{L}_y, \Phi(r)] = 0, \qquad [\hat{L}_z, \Phi(r)] = 0, \qquad [\hat{L}^2, \Phi(r)] = 0. \tag{A3}$$

**A2. Raising and Lowering Operators**. Let the wave-function $\psi_m(\mathbf{r})$ be an eigen-function of the angular momentum operator $\hat{L}_z$. Since $\hat{L}_z$ commutes with $\hat{L}^2$, the same wave-function will also be an eigen-function of $\hat{L}^2$—but not one of $\hat{L}_x$ or $\hat{L}_y$, which do not commute with $\hat{L}_z$. We will show below that $(\hat{L}_x + i\hat{L}_y)\psi_m(\mathbf{r})$ and $(\hat{L}_x - i\hat{L}_y)\psi_m(\mathbf{r})$ are two different eigen-functions of $\hat{L}_z$. In fact, since the entire argument is based on the commutators of the various angular momentum operators (with no reference to position and linear momentum operators $\hat{\mathbf{x}}$ and $\hat{\mathbf{p}}$), we can replace the wave-function $\psi_m(\mathbf{r})$ with the state-vector $|\psi_m\rangle$.

$$\hat{L}_z(\hat{L}_x \pm i\hat{L}_y)|\psi_m\rangle = [(\hat{L}_x\hat{L}_z + i\hbar\hat{L}_y) \pm i(\hat{L}_y\hat{L}_z - i\hbar\hat{L}_x)]|\psi_m\rangle$$

$$= [(\hat{L}_x \pm i\hat{L}_y)\hat{L}_z \pm \hbar(\hat{L}_x \pm i\hat{L}_y)]|\psi_m\rangle$$

$$= (\hat{L}_x \pm i\hat{L}_y)(\hat{L}_z \pm \hbar)|\psi_m\rangle = (m \pm 1)\hbar(\hat{L}_x \pm i\hat{L}_y)|\psi_m\rangle. \tag{A4}$$

In the last line of the above equation, we have set the eigenvalue of $\hat{L}_z$ associated with $|\psi_m\rangle$ equal to $m\hbar$. Since the value of $m$ is as yet unspecified, this choice imposes no restrictions on the state-vector $|\psi_m\rangle$ or on its eigenvalue. It is clear that the operator $\hat{L}_+ = \hat{L}_x + i\hat{L}_y$ raises the eigen-value of $\hat{L}_z$ by one $\hbar$, while $\hat{L}_- = \hat{L}_x - i\hat{L}_y$ lowers the eigen-value by the same amount.

Since $\hat{L}_+$ and $\hat{L}_-$ are Hermitian conjugates of each other, the squared norms of the two states $\hat{L}_+|\psi_m\rangle$ and $\hat{L}_-|\psi_m\rangle$ may be found in the following way:

$$\langle\psi_m|(\hat{L}_x - i\hat{L}_y)(\hat{L}_x + i\hat{L}_y)|\psi_m\rangle = \langle\psi_m|(\hat{L}_x^2 + \hat{L}_y^2 + i[\hat{L}_x, \hat{L}_y])|\psi_m\rangle$$

$$= \langle\psi_m|(\hat{L}^2 - \hat{L}_z^2 - \hbar\hat{L}_z)|\psi_m\rangle = \hbar^2[\ell(\ell+1) - m^2 - m]\langle\psi_m|\psi_m\rangle. \tag{A5a}$$

$$\langle\psi_m|(\hat{L}_x + i\hat{L}_y)(\hat{L}_x - i\hat{L}_y)|\psi_m\rangle = \langle\psi_m|(\hat{L}^2 - \hat{L}_z^2 + \hbar\hat{L}_z)|\psi_m\rangle$$

$$= \hbar^2[\ell(\ell+1) - m^2 + m]\langle\psi_m|\psi_m\rangle. \tag{A5b}$$

Note that the eigenvalue of $\hat{L}^2$ associated with $|\psi_m\rangle$ is written as $\ell(\ell+1)\hbar^2$, independently of the value of $m$. As long as $\ell$ is taken to be arbitrary, this choice of the eigenvalue is not restrictive at all, as $\ell(\ell+1)$ can be any real number $\alpha$. The possible values of $\ell$ will then be $-\tfrac{1}{2} \pm \sqrt{\tfrac{1}{4} + \alpha}$, which, depending on the value of $\alpha$, could be real or imaginary, but $\ell(\ell+1)$ will always turn out to be real and equal to $\alpha$. The important conclusion reached from Eqs.(A5) is that the squared norms, which must be positive, yield the following inequalities:

$$\ell(\ell+1) - m^2 \pm m \geq 0 \;\rightarrow\; (\ell + \tfrac{1}{2})^2 \geq (m \pm \tfrac{1}{2})^2 \;\rightarrow\; |m \pm \tfrac{1}{2}| \leq \ell + \tfrac{1}{2} \;\rightarrow\; -\ell \leq m \leq \ell. \tag{A6}$$

Given that the raising and lowering operators change the value of $m$ in increments of $\pm 1$, one concludes from Eq.(A6) that the correct values of $\ell$ should be either positive integers or positive half-integers. The eigenvalue of $\hat{L}^2$ associated with $|\psi_m\rangle$ may now be determined as follows:

$$\hat{L}^2|\psi_m\rangle = (\hat{L}_x^2 + \hat{L}_y^2 + \hat{L}_z^2)|\psi_m\rangle = [(\hat{L}_x - i\hat{L}_y)(\hat{L}_x + i\hat{L}_y) - i(\hat{L}_x\hat{L}_y - \hat{L}_y\hat{L}_x) + \hat{L}_z^2]|\psi_m\rangle$$

$$= [(\hat{L}_x - i\hat{L}_y)(\hat{L}_x + i\hat{L}_y) + \hbar\hat{L}_z + \hat{L}_z^2]|\psi_m\rangle$$

$$= (\hat{L}_x - i\hat{L}_y)(\hat{L}_x + i\hat{L}_y)|\psi_m\rangle + m(m+1)\hbar^2|\psi_m\rangle. \tag{A7}$$

If $|\psi_m\rangle$ happens to be the eigen-function (or eigenvector) of $\hat{L}_z$ with the largest allowed eigen-value, then $(\hat{L}_x + i\hat{L}_y)|\psi_m\rangle$ will be zero, in which case the eigenvalue of $\hat{L}^2$ will be $\ell(\ell+1)\hbar^2$, where $\ell$ is the largest possible value of $m$. Similarly,

$$\hat{L}^2|\psi_m\rangle = (\hat{L}_x^2 + \hat{L}_y^2 + \hat{L}_z^2)|\psi_m\rangle = [(\hat{L}_x + i\hat{L}_y)(\hat{L}_x - i\hat{L}_y) + i(\hat{L}_x\hat{L}_y - \hat{L}_y\hat{L}_x) + \hat{L}_z^2]|\psi_m\rangle$$

$$= [(\hat{L}_x + i\hat{L}_y)(\hat{L}_x - i\hat{L}_y) - \hbar\hat{L}_z + \hat{L}_z^2]|\psi_m\rangle$$

$$= (\hat{L}_x + i\hat{L}_y)(\hat{L}_x - i\hat{L}_y)|\psi_m\rangle + m(m-1)\hbar^2|\psi_m\rangle. \tag{A8}$$

If $|\psi_m\rangle$ happens to be the eigen-function (or eigenvector) of $\hat{L}_z$ with the smallest allowed eigen-value, then $(\hat{L}_x - i\hat{L}_y)|\psi_m\rangle$ will be zero and the eigenvalue of $\hat{L}^2$ will be $\ell(\ell-1)\hbar^2$, where $\ell$ is the smallest possible value of $m$. In light of the preceding arguments we must have $\ell = -\ell$ and, therefore, once again the eigenvalue of $\hat{L}^2$ is $\ell(\ell+1)\hbar^2$, where $\ell$ is either a positive integer or a positive half-integer. The eigenvalues of $\hat{L}_z$ will then be $m\hbar$, with $m$ assuming all values between $-\ell$ and $\ell$ in increments of 1.

**A3. Angular momentum operator in spherical coordinates**. Considering that the orbital angular momentum operator is $\hat{\boldsymbol{L}} = \boldsymbol{r} \times \hat{\boldsymbol{p}} = -i\hbar\boldsymbol{r} \times \boldsymbol{\nabla}$, we use the expression of the gradient operator in spherical coordinates, namely,

$$\boldsymbol{\nabla}\psi = (\partial_r\psi)\hat{\boldsymbol{r}} + r^{-1}(\partial_\theta\psi)\hat{\boldsymbol{\theta}} + (r\sin\theta)^{-1}(\partial_\phi\psi)\hat{\boldsymbol{\phi}}, \tag{A9}$$

to write $\hat{\boldsymbol{L}}$ in spherical coordinates as follows:

$$\hat{\boldsymbol{L}}\psi = -i\hbar[(\partial_\theta\psi)\hat{\boldsymbol{\phi}} - (\sin\theta)^{-1}(\partial_\phi\psi)\hat{\boldsymbol{\theta}}]. \tag{A10}$$

Now, $\hat{\boldsymbol{\theta}} = \cos\theta\cos\phi\,\hat{\boldsymbol{x}} + \cos\theta\sin\phi\,\hat{\boldsymbol{y}} - \sin\theta\,\hat{\boldsymbol{z}}$ and $\hat{\boldsymbol{\phi}} = -\sin\phi\,\hat{\boldsymbol{x}} + \cos\phi\,\hat{\boldsymbol{y}}$. Consequently,

$$\hat{L}_x = i\hbar(\sin\phi\,\partial_\theta + \cot\theta\cos\phi\,\partial_\phi), \tag{A11a}$$

$$\hat{L}_y = -i\hbar(\cos\phi\,\partial_\theta - \cot\theta\sin\phi\,\partial_\phi), \tag{A11b}$$

$$\hat{L}_z = -i\hbar\partial_\phi. \tag{A11c}$$

It is now easy to verify the commutation relations among the various angular momentum operators using the above expressions of $\hat{L}_x, \hat{L}_y, \hat{L}_z$ in spherical coordinates. Similarly, The squared angular momentum operator $\hat{L}^2$ in spherical coordinates may be obtained as follows:

$$\hat{L}^2 = (\boldsymbol{r} \times \hat{\boldsymbol{p}}) \cdot (\boldsymbol{r} \times \hat{\boldsymbol{p}}) = -\hbar^2(\boldsymbol{r} \times \boldsymbol{\nabla}) \cdot (\boldsymbol{r} \times \boldsymbol{\nabla})$$

$$= -\hbar^2[(y\partial_z - z\partial_y)(y\partial_z - z\partial_y) + (z\partial_x - x\partial_z)(z\partial_x - x\partial_z) + (x\partial_y - y\partial_x)(x\partial_y - y\partial_x)]$$

$$= -\hbar^2[(y^2\partial_z^2 + z^2\partial_y^2 - 2yz\partial_{yz}^2 - y\partial_y - z\partial_z) + (z^2\partial_x^2 + x^2\partial_z^2 - 2xz\partial_{xz}^2 - x\partial_x - z\partial_z)$$
$$+ (x^2\partial_y^2 + y^2\partial_x^2 - 2xy\partial_{xy}^2 - x\partial_x - y\partial_y)]$$
$$= -\hbar^2[(y^2 + z^2)\partial_x^2 + (x^2 + z^2)\partial_y^2 + (x^2 + y^2)\partial_z^2 - 2(xy\partial_{xy}^2 + yz\partial_{yz}^2 + xz\partial_{xz}^2)$$
$$-2(x\partial_x + y\partial_y + z\partial_z)]$$
$$= -\hbar^2[(x^2 + y^2 + z^2)(\partial_x^2 + \partial_y^2 + \partial_z^2) - (x^2\partial_x^2 + y^2\partial_y^2 + z^2\partial_z^2)$$
$$-2(xy\partial_{xy}^2 + yz\partial_{yz}^2 + xz\partial_{xz}^2) - 2(x\partial_x + y\partial_y + z\partial_z)]$$
$$= -\hbar^2[r^2\nabla^2 - (x\partial_x + y\partial_y + z\partial_z)(x\partial_x + y\partial_y + z\partial_z) - (x\partial_x + y\partial_y + z\partial_z)]$$
$$= -\hbar^2[r^2\nabla^2 - (\boldsymbol{r}\cdot\boldsymbol{\nabla})(\boldsymbol{r}\cdot\boldsymbol{\nabla}) - (\boldsymbol{r}\cdot\boldsymbol{\nabla})]. \tag{A12}$$

Using Eq.(A9), the expression of the gradient operator in spherical coordinates, we find
$$(\boldsymbol{r}\cdot\boldsymbol{\nabla})\psi = r\partial_r\psi, \tag{A13a}$$
and
$$(\boldsymbol{r}\cdot\boldsymbol{\nabla})(\boldsymbol{r}\cdot\boldsymbol{\nabla})\psi = (\boldsymbol{r}\cdot\boldsymbol{\nabla})r\partial_r\psi = r\partial_r(r\partial_r\psi) = (r\partial_r + r^2\partial_r^2)\psi. \tag{A13b}$$

We also have
$$r^2\nabla^2\psi = \partial_r(r^2\partial_r\psi) + (\sin\theta)^{-1}\partial_\theta(\sin\theta\,\partial_\theta\psi) + (\sin\theta)^{-2}\partial_\phi^2\psi. \tag{A14}$$

Therefore,
$$\hat{L}^2 = -\hbar^2[(\sin\theta)^{-1}\partial_\theta(\sin\theta\,\partial_\theta) + (\sin\theta)^{-2}\partial_\phi^2]. \tag{A15}$$

The same result, of course, may be obtained directly from $\hat{L}^2 = \hat{L}_x^2 + \hat{L}_y^2 + \hat{L}_z^2$ using Eqs.(A11). If the wave-function for a given energy eigenstate $|\psi\rangle$ is $\psi(r,t) = R(r)Y(\theta,\phi)\exp(-i\mathcal{E}t/\hbar)$, where $\mathcal{E}$ is the energy of the state $|\psi\rangle$, it is clear that the above orbital angular momentum operators operate only on $Y(\theta,\phi)$, thus allowing the radial and temporal functions $R(r)$ and $\exp(-i\mathcal{E}t/\hbar)$ to pass through the operators unchanged.

---

**Digression**: With reference to Eq.(A10), one might be tempted to define $\hat{L}^2 = \hat{L}_\theta^2 + \hat{L}_\phi^2$ where $\hat{L}_\theta = i\hbar(\sin\theta)^{-1}\partial_\phi$ and $\hat{L}_\phi = -i\hbar\partial_\theta$. This turns out to be incorrect, as it is readily seen to lead to $\hat{L}^2 = -\hbar^2[\partial_\theta^2 + (\sin\theta)^{-2}\partial_\phi^2]$, in clear disagreement with Eq.(A15). The correct expression for $\hat{L}^2$ may be derived from Eq.(A10) as follows:

$$\hat{L}^2 = \hat{\boldsymbol{L}}\cdot\hat{\boldsymbol{L}} = -\hbar^2[\hat{\boldsymbol{\phi}}\partial_\theta - \hat{\boldsymbol{\theta}}(\sin\theta)^{-1}\partial_\phi]\cdot[\hat{\boldsymbol{\phi}}\partial_\theta - \hat{\boldsymbol{\theta}}(\sin\theta)^{-1}\partial_\phi] \quad {\scriptstyle -(\sin\theta\,\hat{\boldsymbol{r}} + \cos\theta\,\hat{\boldsymbol{\theta}})}$$
$$= -\hbar^2[(\hat{\boldsymbol{\phi}}\cdot\partial_\theta\hat{\boldsymbol{\phi}})\partial_\theta + \partial_\theta^2 - (\hat{\boldsymbol{\phi}}\cdot\partial_\phi\hat{\boldsymbol{\theta}})(\sin\theta)^{-1}\partial_\phi - (\hat{\boldsymbol{\theta}}\cdot\partial_\phi\hat{\boldsymbol{\phi}})(\sin\theta)^{-1}\partial_\theta$$
$$+ (\hat{\boldsymbol{\theta}}\cdot\partial_\phi\hat{\boldsymbol{\theta}})(\sin\theta)^{-2}\partial_\phi + (\sin\theta)^{-2}\partial_\phi^2]$$
$$= -\hbar^2[\partial_\theta^2 + (\hat{\boldsymbol{\phi}}\cdot\hat{\boldsymbol{r}})(\sin\theta)^{-1}\partial_\phi + \hat{\boldsymbol{\theta}}\cdot(\sin\theta\,\hat{\boldsymbol{r}} + \cos\theta\,\hat{\boldsymbol{\theta}})(\sin\theta)^{-1}\partial_\theta$$
$$+ \hat{\boldsymbol{\theta}}\cdot(\cos\theta\,\hat{\boldsymbol{\phi}})(\sin\theta)^{-2}\partial_\phi + (\sin\theta)^{-2}\partial_\phi^2]$$
$$= -\hbar^2[\partial_\theta^2 + \cot\theta\,\partial_\theta + (\sin\theta)^{-2}\partial_\phi^2]. \tag{A16}$$

The above expression for $\hat{L}^2$ is equivalent to that given by Eq.(A15).

---

**A4. Orbital Angular Momentum Eigenstates**. The functions $Y_{\ell m}(\theta,\phi) = P_{\ell m}(\theta)\exp(im\phi)$ for positive integers $\ell$ and integers $m$, where $-\ell \le m \le \ell$, are the eigenstates of both $\hat{L}^2$ and $\hat{L}_z$. Here $P_{\ell m}(\theta)$ are related (but not identical) to the associated Legendre polynomials $P_\ell^m(\cos\theta)$. The functional form of $P_{\ell m}(\theta)$ may be determined by repeated application of the raising and lowering operators, which, in spherical coordinates, are written

$$\hat{L}_+ = \hat{L}_x + i\hat{L}_y = \hbar \exp(i\phi)(\partial_\theta + i\cot\theta\, \partial_\phi), \tag{A17a}$$

$$\hat{L}_- = \hat{L}_x - i\hat{L}_y = -\hbar \exp(-i\phi)(\partial_\theta - i\cot\theta\, \partial_\phi). \tag{A17b}$$

Applying $\hat{L}_+$ to $Y_{\ell,\ell}(\theta,\phi)$ must yield zero, that is, $P'_{\ell,\ell} - \ell \cot\theta\, P_{\ell,\ell} = 0$. Therefore, $(\ln P_{\ell,\ell})' = (\ell \ln \sin\theta)'$, or $P_{\ell,\ell} = C \sin^\ell\theta$, where $C$ is an arbitrary constant. The normalization condition then yields [Integration by parts using $u' = \sin\theta$ and $v = \sin^{2\ell}(\theta)$ leads to a recursion relation.]

$$2\pi \int_0^\pi P_{\ell,\ell}^2(\theta)\sin\theta\, d\theta = 2\pi C^2 \int_0^\pi \sin^{2\ell+1}(\theta)\, d\theta \overset{\downarrow}{=} 4\pi C^2 \frac{(2\ell)!!}{(2\ell+1)!!} = 1. \tag{A18}$$

Next, we apply $\hat{L}_-$ to $Y_{\ell,\ell}(\theta,\phi) = (-1)^\ell \sqrt{(2\ell+1)!!/4\pi(2\ell)!!}\, \sin^\ell\theta \exp(i\ell\phi)$ to obtain $Y_{\ell,\ell-1}(\theta,\phi)$, as follows: [$(-1)^\ell$ is introduced to satisfy a sign convention.]

$$(\hat{L}_x - i\hat{L}_y)Y_{\ell,\ell}(\theta,\phi) = -\hbar \exp(-i\phi)(\partial_\theta - i\cot\theta\,\partial_\phi)Y_{\ell,\ell}(\theta,\phi)$$

$$= 2(-1)^{\ell-1}\hbar\ell\sqrt{(2\ell+1)!!/4\pi(2\ell)!!}\, \cos\theta \sin^{\ell-1}(\theta)\exp[i(\ell-1)\phi]. \tag{A19}$$

The above function must be normalized by $\hbar\sqrt{\ell(\ell+1) - \ell(\ell-1)} = \hbar\sqrt{2\ell}$, which is the coefficient introduced into the wave-function by the lowering operator $\hat{L}_-$; see Eq.(A5b). We thus find

$$Y_{\ell,\ell-1}(\theta,\phi) = (-1)^{\ell-1}\sqrt{(2\ell+1)!!/4\pi(2\ell-2)!!}\, \cos\theta \sin^{\ell-1}(\theta)\exp[i(\ell-1)\phi]. \tag{A20}$$

In this way, one can calculate all the wave-functions $Y_{\ell m}(\theta,\phi)$ for various integers $\ell$ and $m$.

**A5. Spin Angular Momentum**. The spin angular momentum operator $\hat{\mathbf{S}} = \tfrac{1}{2}\hbar(\sigma_x \hat{x} + \sigma_y \hat{y} + \sigma_z \hat{z})$, [$\hat{x}, \hat{y}, \hat{z}$ are unit-vectors, not operators.] where $\sigma_x = \begin{pmatrix} 0 & 1 \\ 1 & 0 \end{pmatrix}, \sigma_y = \begin{pmatrix} 0 & -i \\ i & 0 \end{pmatrix}, \sigma_z = \begin{pmatrix} 1 & 0 \\ 0 & -1 \end{pmatrix}$ are the Pauli spin matrices, acts on the spinor $\begin{bmatrix} \psi_\uparrow(\mathbf{r},t) \\ \psi_\downarrow(\mathbf{r},t) \end{bmatrix}$. The components of $\hat{\mathbf{S}}$ commute similarly to those of $\hat{\mathbf{L}}$, that is,

$$[S_x, S_y] = i\hbar S_z, \qquad [S_y, S_z] = i\hbar S_x, \qquad [S_z, S_x] = i\hbar S_y. \tag{A21}$$

Therefore, all properties of the orbital angular momentum operator which stem from such commutation relations are equally valid for the spin angular momentum operator. The operator $\hat{S}^2 = \hat{S}_x^2 + \hat{S}_y^2 + \hat{S}_z^2 = \tfrac{3}{4}\hbar^2 \begin{pmatrix} 1 & 0 \\ 0 & 1 \end{pmatrix}$ yields the squared magnitude $s(s+1)\hbar^2$ of the spin angular momentum of arbitrary spinors, considering that the spin quantum number is $s = \tfrac{1}{2}$.

The raising and lowering operators associated with $\hat{\mathbf{S}}$ are $\hat{S}_+ = \hat{S}_x + i\hat{S}_y = \hbar \begin{pmatrix} 0 & 1 \\ 0 & 0 \end{pmatrix}$ and $\hat{S}_- = \hat{S}_x - i\hat{S}_y = \hbar \begin{pmatrix} 0 & 0 \\ 1 & 0 \end{pmatrix}$, while the eigenstates of $\hat{S}_z = \tfrac{1}{2}\hbar\sigma_z$ are $\begin{bmatrix} \psi_\uparrow(\mathbf{r},t) \\ 0 \end{bmatrix}$ and $\begin{bmatrix} 0 \\ \psi_\downarrow(\mathbf{r},t) \end{bmatrix}$, corresponding to $s_z = \pm\tfrac{1}{2}\hbar$, respectively. It is now readily verified that

$$\hat{S}_+\begin{bmatrix}\psi_\uparrow(r,t)\\0\end{bmatrix}=0\quad\text{and}\quad\hat{S}_-\begin{bmatrix}0\\\psi_\downarrow(r,t)\end{bmatrix}=0,$$

whereas

$$\hat{S}_+\begin{bmatrix}0\\\psi_\downarrow(r,t)\end{bmatrix}=\hbar\begin{bmatrix}\psi_\downarrow(r,t)\\0\end{bmatrix}\quad\text{and}\quad\hat{S}_-\begin{bmatrix}\psi_\uparrow(r,t)\\0\end{bmatrix}=\hbar\begin{bmatrix}0\\\psi_\uparrow(r,t)\end{bmatrix}.$$

As expected, the normalization coefficients $\hbar\sqrt{s(s+1)-s_z(s_z\pm 1)}$ are either 0 or $\hbar$.

**A6. Total Angular Momentum**. The total angular momentum operator $\hat{J}=\hat{L}+\hat{S}$ acts on the spinor $\begin{bmatrix}\psi_\uparrow(r,t)\\\psi_\downarrow(r,t)\end{bmatrix}$. The components of $\hat{J}$ obey the same commutation relations as those of $\hat{L}$ and $\hat{S}$, that is,

$$[\hat{J}_x,\hat{J}_y]=i\hbar\hat{J}_z,\qquad [\hat{J}_y,\hat{J}_z]=i\hbar\hat{J}_x,\qquad [\hat{J}_z,\hat{J}_x]=i\hbar\hat{J}_y. \tag{A22}$$

Once again, all properties of the orbital angular momentum operator $\hat{L}$ which stem from such commutation relations are equally valid for the total angular momentum operator $\hat{J}$. In particular, since $\hat{J}^2$ commutes with $\hat{J}_z$, the eigenstates of $\hat{J}^2$ must be eigenstates of $J_z$ as well. We conjecture, therefore, that in a spherically symmetric system the eigenstates of $\hat{J}^2$ are spinors of the type

$$|\psi_{j,m+\frac{1}{2}}(r)\rangle=\begin{bmatrix}\alpha_\uparrow R(r)Y_{\ell,m}(\theta,\phi)\\\alpha_\downarrow R(r)Y_{\ell,m+1}(\theta,\phi)\end{bmatrix}, \tag{A23}$$

where $\alpha_\uparrow$ and $\alpha_\downarrow$ are constant coefficients to be determined shortly. The action of $\hat{J}_z=\hat{L}_z+\hat{S}_z=-i\hbar\partial_\phi+\tfrac{1}{2}\hbar\begin{pmatrix}1&0\\0&-1\end{pmatrix}$ on $|\psi\rangle$ thus yields $(m+\tfrac{1}{2})\hbar|\psi\rangle$, which corresponds to a total angular momentum eigenvalue $j_z=(m+\tfrac{1}{2})\hbar$ along the z-axis. As for the action of $\hat{J}^2$, we write

$$\hat{J}^2=(\hat{L}+\hat{S})\cdot(\hat{L}+\hat{S})=\hat{L}^2+\hat{S}^2+2\hat{L}\cdot\hat{S}=\hat{L}^2+\hat{S}^2+\hbar\begin{pmatrix}\hat{L}_z & \hat{L}_x-i\hat{L}_y\\\hat{L}_x+i\hat{L}_y & -\hat{L}_z\end{pmatrix}. \tag{A24}$$

Recalling that $\hat{L}_x\pm i\hat{L}_y$ are the raising and lowering operators, we will have

$$\hat{J}^2|\psi\rangle=\ell(\ell+1)\hbar^2|\psi\rangle+\tfrac{3}{4}\hbar^2|\psi\rangle+\hbar^2\begin{pmatrix}[\alpha_\uparrow m+\alpha_\downarrow\sqrt{\ell(\ell+1)-m(m+1)}\,]RY_{\ell,m}\\ [\alpha_\uparrow\sqrt{\ell(\ell+1)-m(m+1)}-\alpha_\downarrow(m+1)]RY_{\ell,m+1}\end{pmatrix}. \tag{A25}$$

Therefore, $|\psi\rangle$ will be an eigenstate of $\hat{J}^2$ provided that the following equality holds:

$$m+(\alpha_\downarrow/\alpha_\uparrow)\sqrt{\ell(\ell+1)-m(m+1)}=(\alpha_\uparrow/\alpha_\downarrow)\sqrt{\ell(\ell+1)-m(m+1)}-(m+1). \tag{A26}$$

This quadratic equation in $\alpha_\uparrow/\alpha_\downarrow$ has two solutions, namely,

$$\alpha_\uparrow/\alpha_\downarrow=\sqrt{(\ell+m+1)/(\ell-m)}, \tag{A27a}$$

$$\alpha_\uparrow/\alpha_\downarrow=-\sqrt{(\ell-m)/(\ell+m+1)}. \tag{A27b}$$

Substituting the above values of $\alpha_\uparrow/\alpha_\downarrow$ into Eq.(A25) now yields

$$\hat{J}^2|\psi\rangle=\hbar^2[\ell(\ell+1)+\tfrac{3}{4}+\ell]|\psi\rangle=\hbar^2(\ell+\tfrac{1}{2})(\ell+\tfrac{1}{2}+1)|\psi\rangle, \tag{A28a}$$

$$\hat{J}^2|\psi\rangle = \hbar^2[\ell(\ell+1) + \tfrac{3}{4} - (\ell+1)]|\psi\rangle = \hbar^2(\ell - \tfrac{1}{2})(\ell - \tfrac{1}{2} + 1)|\psi\rangle. \tag{A28b}$$

The eigenvalues of $\hat{J}^2$ are thus seen to be $j(j+1)\hbar^2$, with $j = |\ell \pm \tfrac{1}{2}|$, corresponding to the parallel and anti-parallel orientations of the spin and orbital angular momenta. Normalizing $\alpha_\uparrow$ and $\alpha_\downarrow$ so that $\alpha_\uparrow^2 + \alpha_\downarrow^2 = 1$, we find

$$|\psi_{\ell+\frac{1}{2},m+\frac{1}{2}}(\boldsymbol{r})\rangle = R(r)\begin{bmatrix}\sqrt{(\ell+m+1)/(2\ell+1)}\,Y_{\ell,m}(\theta,\phi)\\ \sqrt{(\ell-m)/(2\ell+1)}\,Y_{\ell,m+1}(\theta,\phi)\end{bmatrix}, \tag{A29a}$$

$$|\psi_{\ell-\frac{1}{2},m+\frac{1}{2}}(\boldsymbol{r})\rangle = R(r)\begin{bmatrix}-\sqrt{(\ell-m)/(2\ell+1)}\,Y_{\ell,m}(\theta,\phi)\\ \sqrt{(\ell+m+1)/(2\ell+1)}\,Y_{\ell,m+1}(\theta,\phi)\end{bmatrix}. \tag{A29b}$$

Acceptable values of $j_z = (m + \tfrac{1}{2})\hbar$ must therefore be in the interval $[-j, j]$. In the preceding equation, when $m = \ell$, one must set $Y_{\ell,m+1}$ equal to zero. Similarly, one sets $Y_{\ell,m}$ equal to zero when $m = -(\ell+1)$. In general, integer values of $\ell \geq 0$ with $j = |\ell \pm \tfrac{1}{2}|$ and $-j \leq (m+\tfrac{1}{2}) \leq j$ always lead to valid spinors.

**A7. Action of $\hat{\boldsymbol{\sigma}} \cdot \boldsymbol{r}/r$ on $|\psi_{j,j_z}(\boldsymbol{r})\rangle$.** In a spherical coordinate system, the operator $\hat{\boldsymbol{\sigma}} \cdot \boldsymbol{r}/r$ may be written as follows:

$$\hat{\boldsymbol{\sigma}} \cdot \boldsymbol{r}/r = \begin{bmatrix} z/r & (x-iy)/r \\ (x+iy)/r & -z/r \end{bmatrix} = \begin{bmatrix} \cos\theta & \sin\theta\exp(-i\phi) \\ \sin\theta\exp(i\phi) & -\cos\theta \end{bmatrix}. \tag{A30}$$

This Hermitian matrix (or operator), being its own transpose conjugate, when multiplied by itself yields the identity matrix. As such, its action on a state-vector $|\psi\rangle$ does not modify the norm $\langle\psi|\psi\rangle$ of the vector. Moreover, upon acting twice on a given vector, it returns the original vector. The action of $\hat{\boldsymbol{\sigma}} \cdot \boldsymbol{r}/r$ on $|\psi_{j,j_z}(\boldsymbol{r})\rangle$, where $j = |\ell \pm \tfrac{1}{2}|$ and $j_z = m + \tfrac{1}{2}$, yields

$$(\hat{\boldsymbol{\sigma}} \cdot \boldsymbol{r}/r)|\psi_{j,j_z}(\boldsymbol{r})\rangle = R(r)\begin{bmatrix}\cos\theta & \sin\theta\exp(-i\phi)\\ \sin\theta\exp(i\phi) & -\cos\theta\end{bmatrix}\begin{bmatrix}\alpha_\uparrow Y_{\ell,m}(\theta,\phi)\\ \alpha_\downarrow Y_{\ell,m+1}(\theta,\phi)\end{bmatrix}$$

$$= R(r)\begin{pmatrix}\alpha_\uparrow[\cos\theta\,P_{\ell,m}(\theta) + (\alpha_\downarrow/\alpha_\uparrow)\sin\theta\,P_{\ell,m+1}(\theta)]\exp(im\phi)\\ \alpha_\downarrow[(\alpha_\uparrow/\alpha_\downarrow)\sin\theta\,P_{\ell,m}(\theta) - \cos\theta\,P_{\ell,m+1}(\theta)]\exp[i(m+1)\phi]\end{pmatrix}. \tag{A31}$$

The resulting spinor is clearly an eigenstate of $\hat{J}_z$ having $j_z = (m+\tfrac{1}{2})\hbar$ and, as will be shown below, is also an eigenstate of $\hat{L}^2$ having either $\ell+1$ or $\ell-1$ for its eigenvalue. To simplify the calculations, we begin by letting $\hat{L}^2 = -\hbar^2[\partial_\theta^2 + \cot\theta\,\partial_\theta + (\sin\theta)^{-2}\partial_\phi^2]$ operate on various functions appearing on the right-hand-side of Eq.(A31), that is,

$$\hat{L}^2\left[\cos\theta\,P_{\ell,m}(\theta)\exp(im\phi)\right]$$
$$= -\hbar^2\{\cos\theta[P''_{\ell,m}(\theta) + \cot\theta\,P'_{\ell,m}(\theta) - m^2(\sin\theta)^{-2}P_{\ell,m}(\theta)]$$
$$\quad - 2\cos\theta\,P_{\ell,m}(\theta) - 2\sin\theta\,P'_{\ell,m}(\theta)\}\exp(im\phi)$$
$$= \hbar^2\left[\ell(\ell+1)\cos\theta\,P_{\ell,m}(\theta) + 2\cos\theta\,P_{\ell,m}(\theta) + 2\sin\theta\,P'_{\ell,m}(\theta)\right]\exp(im\phi). \tag{A32a}$$

$$\hat{L}^2\left[\sin\theta\, P_{\ell,m+1}(\theta)\exp(im\phi)\right]$$
$$= -\hbar^2\{\sin\theta\left[P''_{\ell,m+1}(\theta) + \cot\theta\, P'_{\ell,m+1}(\theta) - m^2(\sin\theta)^{-2}P_{\ell,m+1}(\theta)\right]$$
$$+(\cot\theta\cos\theta - \sin\theta)P_{\ell,m+1}(\theta) + 2\cos\theta\, P'_{\ell,m+1}(\theta)\}\exp(im\phi)$$
$$= \hbar^2\{\ell(\ell+1)\sin\theta\, P_{\ell,m+1}(\theta) - 2(m+\cos^2\theta)(\sin\theta)^{-1}P_{\ell,m+1}(\theta)$$
$$-2\cos\theta\, P'_{\ell,m+1}(\theta)\}\exp(im\phi). \tag{A32b}$$

$$\hat{L}^2\{\sin\theta\, P_{\ell,m}(\theta)\exp[i(m+1)\phi]\}$$
$$= -\hbar^2\{\sin\theta\left[P''_{\ell,m}(\theta) + \cot\theta\, P'_{\ell,m}(\theta) - (m+1)^2(\sin\theta)^{-2}P_{\ell,m}(\theta)\right]$$
$$+(\cot\theta\cos\theta - \sin\theta)P_{\ell,m}(\theta) + 2\cos\theta\, P'_{\ell,m}(\theta)\}\exp[i(m+1)\phi]$$
$$= \hbar^2\{\ell(\ell+1)\sin\theta\, P_{\ell,m}(\theta) + 2(m+\sin^2\theta)(\sin\theta)^{-1}P_{\ell,m}(\theta)$$
$$-2\cos\theta\, P'_{\ell,m}(\theta)\}\exp[i(m+1)\phi]. \tag{A32c}$$

$$\hat{L}^2\{\cos\theta\, P_{\ell,m+1}(\theta)\exp[i(m+1)\phi]\}$$
$$= -\hbar^2\{\cos\theta\left[P''_{\ell,m+1}(\theta) + \cot\theta\, P'_{\ell,m+1}(\theta) - (m+1)^2 P_{\ell,m+1}(\theta)\right]$$
$$-2\cos\theta\, P_{\ell,m+1}(\theta) - 2\sin\theta\, P'_{\ell,m+1}(\theta)\}\exp[i(m+1)\phi]$$
$$= \hbar^2\{\ell(\ell+1)\cos\theta\, P_{\ell,m+1}(\theta) + 2\cos\theta\, P_{\ell,m+1}(\theta)$$
$$+2\sin\theta\, P'_{\ell,m+1}(\theta)\}\exp[i(m+1)\phi]. \tag{A32d}$$

The functions $P'_{\ell,m}(\theta)$ and $P'_{\ell,m+1}(\theta)$ appearing in the above equations may be evaluated by invoking the raising and lowering operators, as follows:

$$\hat{L}_+ Y_{\ell,m}(\theta,\phi) = \hbar\exp(i\phi)\left(\partial_\theta + i\cot\theta\,\partial_\phi\right)P_{\ell,m}(\theta)\exp(im\phi)$$
$$= \hbar\left[P'_{\ell,m}(\theta) - m\cot\theta\, P_{\ell,m}(\theta)\right]\exp[i(m+1)\phi]$$
$$= \hbar\sqrt{\ell(\ell+1) - m(m+1)}\, P_{\ell,m+1}(\theta)\exp[i(m+1)\phi]$$
$$\rightarrow\quad P'_{\ell,m}(\theta) = m\cot\theta\, P_{\ell,m}(\theta) + \sqrt{(\ell-m)(\ell+m+1)}\, P_{\ell,m+1}(\theta). \tag{A33a}$$

$$\hat{L}_- Y_{\ell,m+1}(\theta,\phi) = -\hbar\exp(-i\phi)\left(\partial_\theta - i\cot\theta\,\partial_\phi\right)P_{\ell,m+1}(\theta)\exp[i(m+1)\phi]$$
$$= -\hbar\left[P'_{\ell,m+1}(\theta) + (m+1)\cot\theta\, P_{\ell,m+1}(\theta)\right]\exp(im\phi)$$
$$= \hbar\sqrt{\ell(\ell+1) - m(m+1)}\, P_{\ell,m}(\theta)\exp(im\phi)$$
$$\rightarrow\quad P'_{\ell,m+1}(\theta) = -(m+1)\cot\theta\, P_{\ell,m+1}(\theta) - \sqrt{(\ell-m)(\ell+m+1)}\, P_{\ell,m}(\theta). \tag{A33b}$$

Substitution from Eqs.(A33) into Eqs.(A32) now yields

$$\hat{L}^2\{[\cos\theta\, P_{\ell,m}(\theta) + (\alpha_\downarrow/\alpha_\uparrow)\sin\theta\, P_{\ell,m+1}(\theta)]\exp(im\phi)\}$$

$$= \hbar^2 \left\{ \left[ \ell(\ell+1) + 2(m+1) + 2(\alpha_\downarrow/\alpha_\uparrow)\sqrt{(\ell-m)(\ell+m+1)} \right] \cos\theta\, P_{\ell,m}(\theta) \right.$$

$$\left. + (\alpha_\downarrow/\alpha_\uparrow)\left[ \ell(\ell+1) - 2m + 2(\alpha_\uparrow/\alpha_\downarrow)\sqrt{(\ell-m)(\ell+m+1)} \right] \sin\theta\, P_{\ell,m+1}(\theta) \right\} \exp(im\phi)$$

$$= \begin{cases} (\ell+1)(\ell+2)\hbar^2 \left[\cos\theta\, P_{\ell,m}(\theta) + (\alpha_\downarrow/\alpha_\uparrow)\sin\theta\, P_{\ell,m+1}(\theta)\right]\exp(im\phi); & \text{if } \frac{\alpha_\uparrow}{\alpha_\downarrow} = \sqrt{\frac{\ell+m+1}{\ell-m}}, \\ (\ell-1)\ell\hbar^2 \left[\cos\theta\, P_{\ell,m}(\theta) + (\alpha_\downarrow/\alpha_\uparrow)\sin\theta\, P_{\ell,m+1}(\theta)\right]\exp(im\phi); & \text{if } \frac{\alpha_\uparrow}{\alpha_\downarrow} = -\sqrt{\frac{\ell-m}{\ell+m+1}}. \end{cases} \quad \text{(A34a)}$$

$$\hat{L}^2\{[(\alpha_\uparrow/\alpha_\downarrow)\sin\theta\, P_{\ell,m}(\theta) - \cos\theta\, P_{\ell,m+1}(\theta)]\exp[i(m+1)\phi]\}$$

$$= \hbar^2 \left\{ (\alpha_\uparrow/\alpha_\downarrow)\left[\ell(\ell+1) + 2(m+1) + 2(\alpha_\downarrow/\alpha_\uparrow)\sqrt{(\ell-m)(\ell+m+1)}\right]\sin\theta\, P_{\ell,m}(\theta) \right.$$

$$\left. - \left[\ell(\ell+1) - 2m + 2(\alpha_\uparrow/\alpha_\downarrow)\sqrt{(\ell-m)(\ell+m+1)}\right]\cos\theta\, P_{\ell,m+1}(\theta) \right\}\exp[i(m+1)\phi]$$

$$= \begin{cases} (\ell+1)(\ell+2)\hbar^2\left[(\alpha_\uparrow/\alpha_\downarrow)\sin\theta\, P_{\ell,m}(\theta) - \cos\theta\, P_{\ell,m+1}(\theta)\right]\exp[i(m+1)\phi]; & \text{if } \frac{\alpha_\uparrow}{\alpha_\downarrow} = \sqrt{\frac{\ell+m+1}{\ell-m}}, \\ (\ell-1)\ell\hbar^2\left[(\alpha_\uparrow/\alpha_\downarrow)\sin\theta\, P_{\ell,m}(\theta) - \cos\theta\, P_{\ell,m+1}(\theta)\right]\exp[i(m+1)\phi]; & \text{if } \frac{\alpha_\uparrow}{\alpha_\downarrow} = -\sqrt{\frac{\ell-m}{\ell+m+1}}. \end{cases}$$
(A34b)

Finally, we examine the action of the operator $2\hat{\boldsymbol{L}} \cdot \hat{\boldsymbol{S}}$ on the spinor of Eq.(A31). Considering that

$$2\hat{\boldsymbol{L}} \cdot \hat{\boldsymbol{S}} = \hbar \begin{pmatrix} \hat{L}_z & \hat{L}_- \\ \hat{L}_+ & -\hat{L}_z \end{pmatrix} = \hbar^2 \begin{bmatrix} -i\partial_\phi & -\exp(-i\phi)(\partial_\theta - i\cot\theta\, \partial_\phi) \\ \exp(i\phi)(\partial_\theta + i\cot\theta\, \partial_\phi) & i\partial_\phi \end{bmatrix}, \quad \text{(A35)}$$

we proceed to determine the upper and lower elements of the resulting spinor as follows:

**Upper element**: $-i\partial_\phi[\alpha_\uparrow \cos\theta\, P_{\ell,m}(\theta) + \alpha_\downarrow \sin\theta\, P_{\ell,m+1}(\theta)]\exp(im\phi)$

$$- \exp(-i\phi)(\partial_\theta - i\cot\theta\, \partial_\phi)[\alpha_\uparrow \sin\theta\, P_{\ell,m}(\theta) - \alpha_\downarrow \cos\theta\, P_{\ell,m+1}(\theta)]\exp[i(m+1)\phi]$$

$$= [\alpha_\uparrow m\cos\theta\, P_{\ell,m}(\theta) + \alpha_\downarrow m\sin\theta\, P_{\ell,m+1}(\theta)]\exp(im\phi)$$

$$- \{\alpha_\uparrow \cos\theta\, P_{\ell,m}(\theta) + \alpha_\uparrow \sin\theta\, P'_{\ell,m}(\theta) + \alpha_\downarrow \sin\theta\, P_{\ell,m+1}(\theta) - \alpha_\downarrow \cos\theta\, P'_{\ell,m+1}(\theta)$$

$$+ (m+1)\cot\theta\left[\alpha_\uparrow \sin\theta\, P_{\ell,m}(\theta) - \alpha_\downarrow \cos\theta\, P_{\ell,m+1}(\theta)\right]\}\exp(im\phi)$$

$$= -\left\{[(m+2) + (\alpha_\downarrow/\alpha_\uparrow)\sqrt{(\ell-m)(\ell+m+1)}]\alpha_\uparrow \cos\theta\, P_{\ell,m}(\theta)\right.$$

$$\left. + [(\alpha_\uparrow/\alpha_\downarrow)\sqrt{(\ell-m)(\ell+m+1)} - (m-1)]\alpha_\downarrow \sin\theta\, P_{\ell,m+1}(\theta)\right\}\exp(im\phi)$$

$$= \begin{cases} -(\ell+2)[\alpha_\uparrow \cos\theta\, P_{\ell,m}(\theta) + \alpha_\downarrow \sin\theta\, P_{\ell,m+1}(\theta)]\exp(im\phi); & \text{if } \frac{\alpha_\uparrow}{\alpha_\downarrow} = \sqrt{\frac{\ell+m+1}{\ell-m}}, \\ (\ell-1)[\alpha_\uparrow \cos\theta\, P_{\ell,m}(\theta) + \alpha_\downarrow \sin\theta\, P_{\ell,m+1}(\theta)]\exp(im\phi); & \text{if } \frac{\alpha_\uparrow}{\alpha_\downarrow} = -\sqrt{\frac{\ell-m}{\ell+m+1}}. \end{cases} \quad \text{(A36a)}$$

**Lower element**: $\exp(i\phi)(\partial_\theta + i\cot\theta\, \partial_\phi)[\alpha_\uparrow \cos\theta\, P_{\ell,m}(\theta) + \alpha_\downarrow \sin\theta\, P_{\ell,m+1}(\theta)]\exp(im\phi)$

$$+ i\partial_\phi[\alpha_\uparrow \sin\theta\, P_{\ell,m}(\theta) - \alpha_\downarrow \cos\theta\, P_{\ell,m+1}(\theta)]\exp[i(m+1)\phi]$$

$$= \{-\alpha_\uparrow \sin\theta\, P_{\ell,m}(\theta) + \alpha_\uparrow \cos\theta\, P'_{\ell,m}(\theta) + \alpha_\downarrow \cos\theta\, P_{\ell,m+1}(\theta) + \alpha_\downarrow \sin\theta\, P'_{\ell,m+1}(\theta)$$

$$-m\cot\theta\, [\alpha_\uparrow \cos\theta\, P_{\ell,m}(\theta) + \alpha_\downarrow \sin\theta\, P_{\ell,m+1}(\theta)]$$

$$-\alpha_\uparrow(m+1)\sin\theta\, P_{\ell,m}(\theta) + \alpha_\downarrow(m+1)\cos\theta\, P_{\ell,m+1}(\theta)\}\exp[i(m+1)\phi]$$

$$= -\{[(m+2) + (\alpha_\downarrow/\alpha_\uparrow)\sqrt{(\ell-m)(\ell+m+1)}]\alpha_\uparrow \sin\theta\, P_{\ell,m}(\theta)$$

$$+ [(m-1) - (\alpha_\uparrow/\alpha_\downarrow)\sqrt{(\ell-m)(\ell+m+1)}]\alpha_\downarrow \cos\theta\, P_{\ell,m+1}(\theta)\}\exp[i(m+1)\phi]$$

$$= \begin{cases} -(\ell+2)[\alpha_\uparrow \sin\theta\, P_{\ell,m}(\theta) - \alpha_\downarrow \cos\theta\, P_{\ell,m+1}(\theta)]\exp[i(m+1)\phi]; & \text{if } \dfrac{\alpha_\uparrow}{\alpha_\downarrow} = \sqrt{\dfrac{\ell+m+1}{\ell-m}}, \\ (\ell-1)[\alpha_\uparrow \sin\theta\, P_{\ell,m}(\theta) - \alpha_\downarrow \cos\theta\, P_{\ell,m+1}(\theta)]\exp[i(m+1)\phi]; & \text{if } \dfrac{\alpha_\uparrow}{\alpha_\downarrow} = -\sqrt{\dfrac{\ell-m}{\ell+m+1}}. \end{cases} \quad \text{(A36b)}$$

It is seen that the action of $\hat{\boldsymbol{\sigma}} \cdot \boldsymbol{r}/r$ on the spinor $\begin{bmatrix} \alpha_\uparrow Y_{\ell,m} \\ \alpha_\downarrow Y_{\ell,m+1} \end{bmatrix}$ produces a new spinor whose $\ell$ value either increases by 1 or decreases by 1, while its $m$ value remains intact. The total angular momentum of the resulting spinor is obtained by applying the operator $\hat{J}^2 = \hat{L}^2 + \hat{S}^2 + 2\hat{\boldsymbol{L}}\cdot\hat{\boldsymbol{S}}$, whose eigenvalues are readily seen to be

$$\begin{cases} (\ell+1)(\ell+2)\hbar^2 + \tfrac{3}{4}\hbar^2 - (\ell+2)\hbar^2 = \left(\ell+\tfrac{1}{2}\right)\left(\ell+\tfrac{3}{2}\right)\hbar^2; & \text{if } \dfrac{\alpha_\uparrow}{\alpha_\downarrow} = \sqrt{\dfrac{\ell+m+1}{\ell-m}}, \\ (\ell-1)\ell\hbar^2 + \tfrac{3}{4}\hbar^2 + (\ell-1)\hbar^2 = \left(\ell-\tfrac{1}{2}\right)\left(\ell+\tfrac{1}{2}\right)\hbar^2; & \text{if } \dfrac{\alpha_\uparrow}{\alpha_\downarrow} = -\sqrt{\dfrac{\ell-m}{\ell+m+1}}. \end{cases} \quad \text{(A37)}$$

Consequently, the action of $\hat{\boldsymbol{\sigma}} \cdot \boldsymbol{r}/r$ on $|\psi_{j,j_z}(\boldsymbol{r})\rangle$, while changing the value of $\ell$ by $\pm 1$, does *not* alter the values of $j$ and $j_z$. If the spin and orbital momenta of the system initially happen to be parallel to each other, the action of $\hat{\boldsymbol{\sigma}} \cdot \boldsymbol{r}/r$ causes them to become antiparallel, while raising the value of $\ell$ by 1. Alternatively, if the spin and orbital momenta happen to be antiparallel at first, then the action of $\hat{\boldsymbol{\sigma}} \cdot \boldsymbol{r}/r$ brings them into alignment while lowering the value of $\ell$ by 1. It is now obvious that applying the same operator twice must return the system to its initial state. Note that the adopted sign convention for $Y_{\ell,m}$ dictates that $(\hat{\boldsymbol{\sigma}} \cdot \boldsymbol{r}/r)|\psi_{\ell,m+\frac{1}{2}}(\boldsymbol{r})\rangle = -|\psi_{\ell\pm1,m+\frac{1}{2}}(\boldsymbol{r})\rangle$.